\documentclass[letterpaper, twocolumn,10pt]{article}
\usepackage{usenix}
\usepackage{hyperref}
\usepackage{url}
\usepackage{listings}
\usepackage{pifont}
\usepackage{tikz}
\usepackage{subfig}
\usepackage{multirow}
\usepackage{pgfplots}
\usepackage{booktabs}
\usepackage{cleveref}
\usepackage{dcolumn}
\usepackage{cancel}
\usepackage{ifmtarg}
\usepackage{verbatim}
\usepackage{graphicx}
\usepackage{chngpage}
\usepackage{xspace}
\usepackage{algorithm}
\usepackage[noend]{algpseudocode}
\usepackage{balance}
\usepackage{amsmath}
\usepackage{amssymb}
\usepackage{amsthm}
\usepackage{times}
\usepackage{fullpage}
\usepackage{footnote}
\usepackage{balance}
\usepackage[symbol]{footmisc}
\usepackage{threeparttable}
\usepackage[small,compact]{titlesec}
\usepackage{microtype}
\usepackage{tabularx}
\usepackage{titling}

\newcommand{\Sys}{FlexFlow\xspace}

\newcommand{\er}[1]{\mbox{\rm\em #1}}
\newcommand{\inception}{Inception-v3\xspace}
\newcommand{\resnet}{ResNet-101\xspace}

\posttitle{\par\end{center}}
\postdate{\par\end{center}\vspace{-1.1cm}}

\begin{document}

\title{\bf Beyond Data and Model Parallelism for Deep Neural Networks}
\author{Zhihao Jia \and Matei Zaharia \\ Stanford University \and Alex Aiken}
\date{}
\maketitle

\thispagestyle{empty}

\begin{abstract}


The computational requirements for training deep neural networks (DNNs) have grown to the point that it is now standard practice to parallelize training.
Existing deep learning systems commonly use data or model parallelism, but unfortunately, these strategies often result in suboptimal parallelization performance.

In this paper, we define a more comprehensive search space of parallelization strategies for DNNs called SOAP, which includes strategies to parallelize a DNN in the Sample, Operation, Attribute, and Parameter dimensions.
We also propose \Sys, a deep learning framework that uses guided randomized search of the SOAP space to find a fast parallelization strategy for a specific parallel machine.
To accelerate this search, \Sys introduces a novel execution simulator that can accurately predict a parallelization strategy's performance and is three orders of magnitude faster than prior approaches that have to execute each strategy.
We evaluate \Sys with six real-world DNN benchmarks on two GPU clusters and show that \Sys can increase training throughput by up to 3.8$\times$ over state-of-the-art approaches, even when including its search time, and also improves scalability.
\end{abstract}

\section{Introduction}
\label{sec:intro}
Over the past few years, deep neural networks (DNNs) have driven advances in many practical problems, such as image classification~\cite{alexnet, vgg}, speech recognition~\cite{speech1, speech2}, machine translation~\cite{GNMT, NMT2}, and game playing~\cite{game}. 
Because sophisticated DNN models~\cite{densenet, inception} and larger training datasets~\cite{imagenet2009, MSGBK} have increased the computational requirements to train DNN models,
it is now standard practice to parallelize training across distributed heterogeneous clusters~\cite{tensorflow, distbelief}.

Although DNN applications and the clusters used to parallelize them are increasingly complex, the strategies used by today's deep learning systems (e.g., TensorFlow~\cite{tensorflow}, PyTorch~\cite{pytorch}, Caffe2~\cite{caffe2}, MXNet~\cite{mxnet}) to parallelize training remain simple.
The most common parallelization technique is {\em data parallelism}~\cite{alexnet}, which places a replica of the entire neural network on each device, so that each device processes a subset of the training data and synchronizes network parameters in different replicas at the end of an iteration.
Data parallelism is efficient for compute-intensive DNN operations with a few trainable parameters (e.g., convolution) but achieves suboptimal parallelization performance for operations with a large number of parameters (e.g., matrix-multiplication).
Another common parallelization strategy is {\em model parallelism}~\cite{distbelief}, which assigns disjoint subsets of a neural network each to a dedicated device.
Model parallelism eliminates parameter synchronization between devices but requires data transfers between operations and disallows parallelism within an operation.

Previous work~\cite{OWT, GNMT} has proposed expert-designed strategies that manually optimize parallelization based on human experts' domain knowledge and intuitions.
For example, \cite{OWT} uses data parallelism for convolutional and pooling layers and switches to model parallelism for fully-connected layers to accelerate training convolutional neural networks. 
Expert-designed strategies achieve improved performance compared to data and model parallelism but still result in suboptimal behaviors. 
Section~\ref{sec:eval} shows that we are able to find parallelization strategies that are up to 2.3$\times$ faster than expert-designed strategies.

In addition to these manually designed parallelization strategies, recent work has proposed automated frameworks~\cite{DevicePlace, OptCNN} for finding efficient parallelization strategies in a limited search space. 
For example, REINFORCE~\cite{DevicePlace} uses a reinforcement learning model to learn efficient operation assignments for model parallelism by running diverse strategies on real devices.
As another example, OptCNN~\cite{OptCNN} is designed for parallelizing DNNs with linear computation graphs (e.g., AlexNet~\cite{alexnet}, VGG~\cite{vgg}) and automatically finds strategies that exploit parallelism within each DNN operation.
Existing automated frameworks only explore either parallelism across different operations (e.g., REINFORCE) or parallelism within a single operation (e.g., OptCNN) and therefore miss faster strategies that use parallelism in both dimensions.
We show that exploring a broader search space discovers parallelization strategies 1.2-3.8$\times$ faster than existing automated frameworks (see Section~\ref{sec:eval}).


In this paper, we present \Sys, a deep learning framework that automatically finds fast parallelization strategies over a significantly broader search space than previous systems.
To formalize the problem, we first define the {\em SOAP} (Sample-Operation-Attribute-Parameter) search space of parallelization strategies for DNNs. 
The {\em operation} dimension describes how different operations in a DNN are parallelized.
In addition, for a single DNN operation, the {\em sample} and {\em parameter} dimensions indicate how training samples and model parameters are distributed across devices.
Finally, the {\em attribute} dimension defines how different attributes within a sample are partitioned.
Compared to existing systems that parallelize DNNs in a subset of SOAP dimensions, \Sys considers parallelizing DNNs in all these dimensions and therefore
defines a more comprehensive search space that includes existing approaches as special cases.


A key challenge with the much larger SOAP search space is effectively evaluating candidate parallelization strategies to find an efficient one. Prior work such as REINFORCE~\cite{DevicePlace} relies on executing each parallelization strategy on the hardware for one iteration to measure its execution time. Unfortunately, this approach becomes prohibitively expensive with the multiple orders of magnitude larger SOAP search space.



To address this problem, \Sys introduces a novel {\em execution simulator} that is accurate for predicting the performance of a parallelization strategy and is three orders of magnitude faster than profiling real executions.
The challenge in designing the simulator is how to accurately estimate the execution time of different DNN operators (e.g., convolution and matrix multiplication), which scale non-linearly in a hardware-dependent way with the data.
The \Sys simulator relies on the following two facts: (1) many DNN models use a small number of \emph{distinct} operators (e.g., a neural machine translation model~\cite{GNMT} with hundreds of operators only uses four distinct operators); and (2) the execution time of each DNN operator is typically low-variance and largely independent of the contents of the input data.

The \Sys simulator measures the execution time of an operation once for each input size and uses the measured time to predict all operations with the same type, which only takes tens of milliseconds.
These estimates are then used to predict the performance of a wide variety of parallelization strategies.
In addition, the execution simulator uses a {\em delta simulation algorithm} that simulates a new strategy using incremental updates to previous simulations. 
Compared to existing approaches~\cite{DevicePlace, DevicePlace2} that measure the performance from real executions, our approach has two advantages.
First, the \Sys simulator is much faster.
As a comparison, REINFORCE~\cite{DevicePlace} 
requires 12-27 hours to find an efficient operation assignment for model parallelism on 4 GPUs, while the \Sys simulator enables exploring a more comprehensive search space and finding better parallelization strategies (with 3.4-3.8x higher throughput than REINFORCE) in 14-40 seconds. 
Furthermore, REINFORCE uses 160 compute nodes (with 4 GPUs on each node) to find an efficient strategy in tens of hours, while our experiments use only a single compute node for the simulator.

The execution simulator also achieves high accuracy for predicting parallelization performance.
We evaluate the simulator with six real-world DNNs on two different GPU clusters and show that, for all the measured executions, the relative difference between the real and simulated execution time is less than 30\%.
Most importantly for the search, we test different strategies for a given DNN application and show that their simulated execution time preserves real execution time ordering.


Using the execution simulator as an oracle, the \Sys {\em { }execution optimizer} uses a general Markov Chain Monte Carlo (MCMC) search algorithm (other search strategies could also be used) to explore the SOAP search space and iteratively propose candidate strategies based on the simulated performance of previous candidates. When the search procedure is finished, the execution optimizer returns the best strategy it has discovered.


We evaluate \Sys on a variety of real-world DNN benchmarks including image classification~\cite{alexnet, resnet, inception}, text classification~\cite{RNNTC}, language modeling~\cite{RNNLM}, and neural machine translation~\cite{GNMT}.
Compared to data/model parallelism and expert-designed parallelization strategies~\cite{OWT, GNMT}, \Sys increases training throughput by up to 3.3$\times$, reduces communication costs by up to 5$\times$, and achieves significantly better scaling.
In addition, \Sys also outperforms the strategies found by REINFORCE by 3.4-3.8$\times$ on the same hardware configuration evaluated in REINFORCE, and outperforms OptCNN by 1.2-1.6$\times$, by supporting a broader search space.


To summarize, our contributions are:
\begin{itemize}
\vspace{-2mm}
\item We define the SOAP search space for parallelizing DNN applications, which includes strategies that parallelize in any combination of the sample, operation, attribute, and parameter dimensions.
\vspace{-2mm}
\item We show that under reasonable assumptions it is possible to reliably predict the execution time of parallelized DNNs using a simulator that is three orders of magnitude faster than actually running the DNNs directly on the hardware. 
\vspace{-2mm}
\item We describe \Sys, a deep learning framework that can search for and execute strategies from the entire SOAP space to accelerate DNN training.
\vspace{-2mm}
\item 
We show that \Sys can increase training throughput by up to 3.8$\times$ over state-of-the-art parallelization approaches while improving scalability.
\end{itemize}

\section{Related Work}
\label{sec:related}

{\bf Data and model parallelism} have been widely used by existing deep learning systems (e.g., TensorFlow~\cite{tensorflow}, Caffe2~\cite{caffe2}, and PyTorch~\cite{pytorch}) to distribute the training process across devices. 
Data parallelism~\cite{alexnet} keeps a copy of an entire DNN on each device, which is inefficient for operations with a large number of parameters (e.g., densely-connected layers) and becomes a scalability bottleneck in large scale distributed training.
Model parallelism~\cite{NMT2, distbelief} splits a DNN into disjoint subsets and trains each subset on a dedicated device, which reduces communication costs for synchronizing network parameters in a DNN but exposes limited parallelism.

{\bf Expert-designed parallelization strategies} manually optimize parallelization for specific DNNs by using experts' domain knowledge and experience. 
For example, ~\cite{OWT} introduces ``one weird trick'' that uses data parallelism for convolutional and pooling layers and switches to model parallelism for densely-connected layers to accelerate convolutional neural networks.
To parallelize recurrent neural networks, ~\cite{GNMT} uses data parallelism that replicates the entire DNN on each compute node and switches to model parallelism for intra-node parallelization.
Although these expert-designed parallelization strategies achieve performance improvement over data and model parallelism, they are suboptimal.
We use these expert-designed strategies as baselines in our experiments and show that \Sys can further improve training performance by up to 2.3$\times$.

{\bf Automated frameworks} have been proposed for finding efficient parallelization strategies in a limited search space. 
REINFORCE~\cite{DevicePlace} uses reinforcement learning to find efficient device placement for model parallelism.
OptCNN~\cite{OptCNN} is designed for parallelizing DNNs with linear computation graphs and automatically finds efficient strategies that exploit parallelism within an operation.


\begin{figure}[t]
\centering
\includegraphics[scale=0.40]{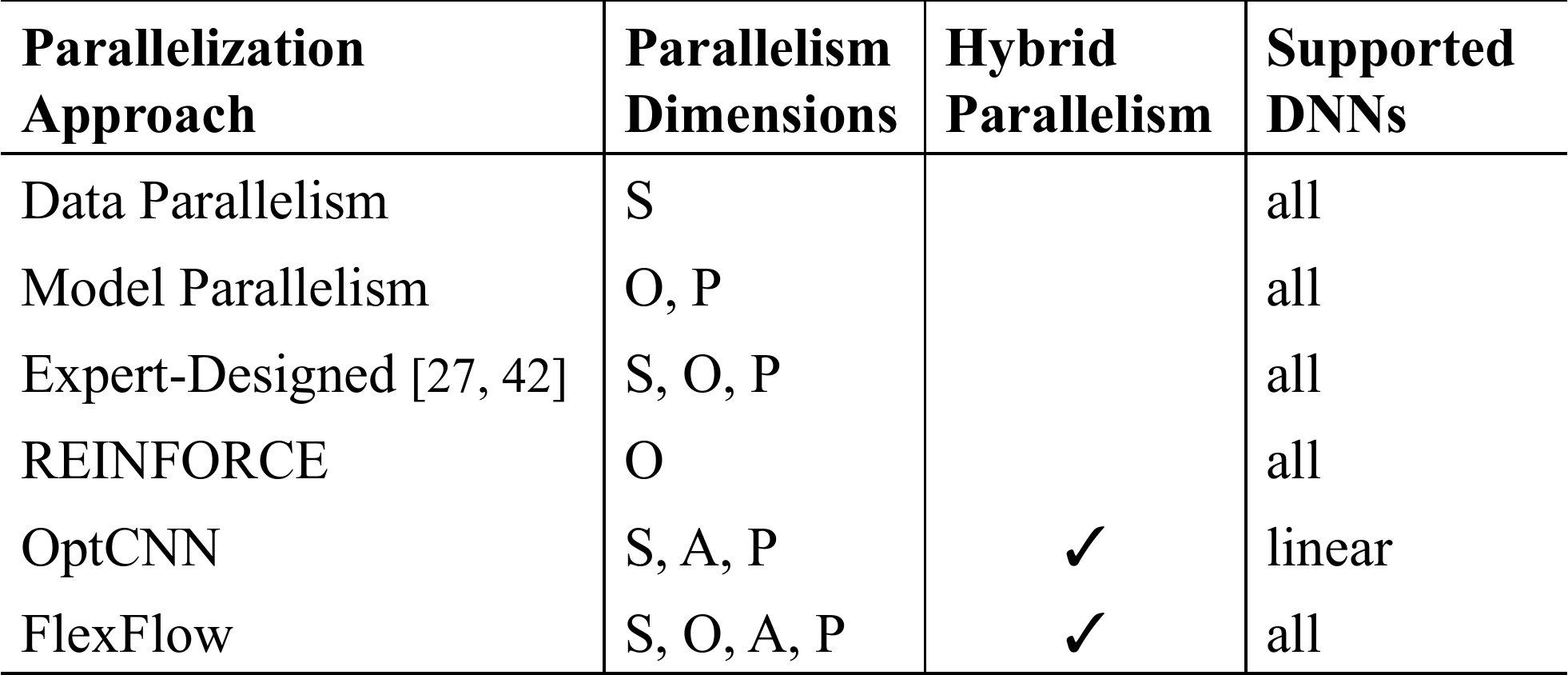}
\vspace{-6mm}
\caption{The parallelism dimensions explored by different approaches.
S, O, A, and P indicate parallelism in the Sample, Operation, Attribute, and Parameter dimensions (see Section~\ref{sec:problem}).
Hybrid parallelism shows if an approach supports parallelizing an operation in a combination of the sample, attribute, and parameter dimensions (see FIgure~\ref{fig:example_parallelization}).
OptCNN is designed for DNNs with linear computation graphs.}
\label{fig:table_comparison}
\vspace{-3mm}
\end{figure}

Figure~\ref{fig:table_comparison} summarizes the parallelism dimensions explored by existing approaches.
Data parallelism uses the sample dimension to parallelize the training process, while model parallelism exploits the parameter and operation dimensions.
Expert-designed strategies~\cite{OWT, GNMT} exploit parallelism in the sample or parameter dimension to parallelize an operation but do not support hybrid parallelism that uses a combination of the sample, attribute, and parameter dimensions to parallelize an operation (see Figure~\ref{fig:example_parallelization}).
Compared to these manually designed strategies, \Sys considers more sophisticated, and often more efficient, strategies to parallelize a single operation.
In addition, compared to existing automated frameworks~\cite{DevicePlace, OptCNN}, \Sys explores a significantly broader search space and is able to find strategies that are up to 3.8$\times$ faster.


{\bf Graph-based cluster schedulers.} Previous work~\cite{quincy, firmament} has proposed cluster schedulers that schedule cluster-wide tasks by using graph-based algorithms. 
For example, Quincy~\cite{quincy} maps task scheduling to a flow network and uses a min-cost max-flow (MCMF) algorithm to find efficient task placement. 
Firmament~\cite{firmament} generalizes Quincy by employing multiple MCMF optimization algorithms to reduce task placement latencies.
Existing graph-based schedulers optimize task placement by assuming a fixed task graph. 
However, \Sys solves a different problem that requires jointly optimizing how to partition an operation into tasks by exploiting parallelism in the SOAP dimensions and how to assign tasks to devices.

\section{Overview}
\label{sec:arch}
In this section, we compare the \Sys programming interface with other frameworks in Section~\ref{subsec:interface}, provide a general overview of \Sys in Section~\ref{subsec:arch}, and discuss the limitations of our approach in Section~\ref{subsec:limitations}.


\subsection{Programming Interface}
\label{subsec:interface}
Similar to existing deep learning systems~\cite{tensorflow, pytorch, caffe2}, \Sys uses an {\em operator graph} $\mathcal{G}$ to describe all operations and state in a DNN. 
Each node $o_i \in \mathcal{G}$ is an operation (e.g., matrix multiplication, convolution, etc.), and each edge $(o_i, o_j) \in \mathcal{G}$ is a tensor (i.e., a $n$-dimensional array) that is an output of $o_i$ and an input of $o_j$.

As far as we know, most deep learning systems (e.g., TensorFlow~\cite{tensorflow}, PyTorch~\cite{pytorch}, and Caffe2~\cite{caffe2}) use data parallelism as the default parallelization strategy and support model parallelism as an alternative by allowing users to manually specify the device placement for each operation. 

In contrast, \Sys takes a {\em device topology} $\mathcal{D} = (\mathcal{D}_N, \mathcal{D}_E)$ describing all available hardware devices and their interconnections, as shown in Figure~\ref{fig:overview}.
Each node $d_i \in \mathcal{D}_N$ represents a device (e.g., a CPU or a GPU), and each edge $(d_i, d_j) \in \mathcal{D}_E$ is a hardware connection (e.g., a NVLink, a PCI-e, or a network link) between device $d_i$ and $d_j$. The edges are labeled with the bandwidth and latency of the connection.

\Sys automatically finds a parallelization strategy for an operator graph and a device topology. Compared to existing frameworks, \Sys has two advantages:

{\bf Programmability.} For DNN applications with complex operator graphs running on clusters with deep device topologies, it is difficult for application developers, even domain experts, to manually design efficient operation assignments.
\Sys takes the responsibility for finding efficient parallelization strategies and provides a more productive programming interface.

{\bf Portability.} 
A parallelization strategy fine-tuned for one cluster may behave poorly on other clusters. 
\Sys's search method automatically selects an efficient strategy for each hardware configuration, without requiring application changes.

\subsection{\Sys Architecture}
\label{subsec:arch}
\begin{figure}[t]
\vspace{-3mm}
\centering
\includegraphics[scale=0.44]{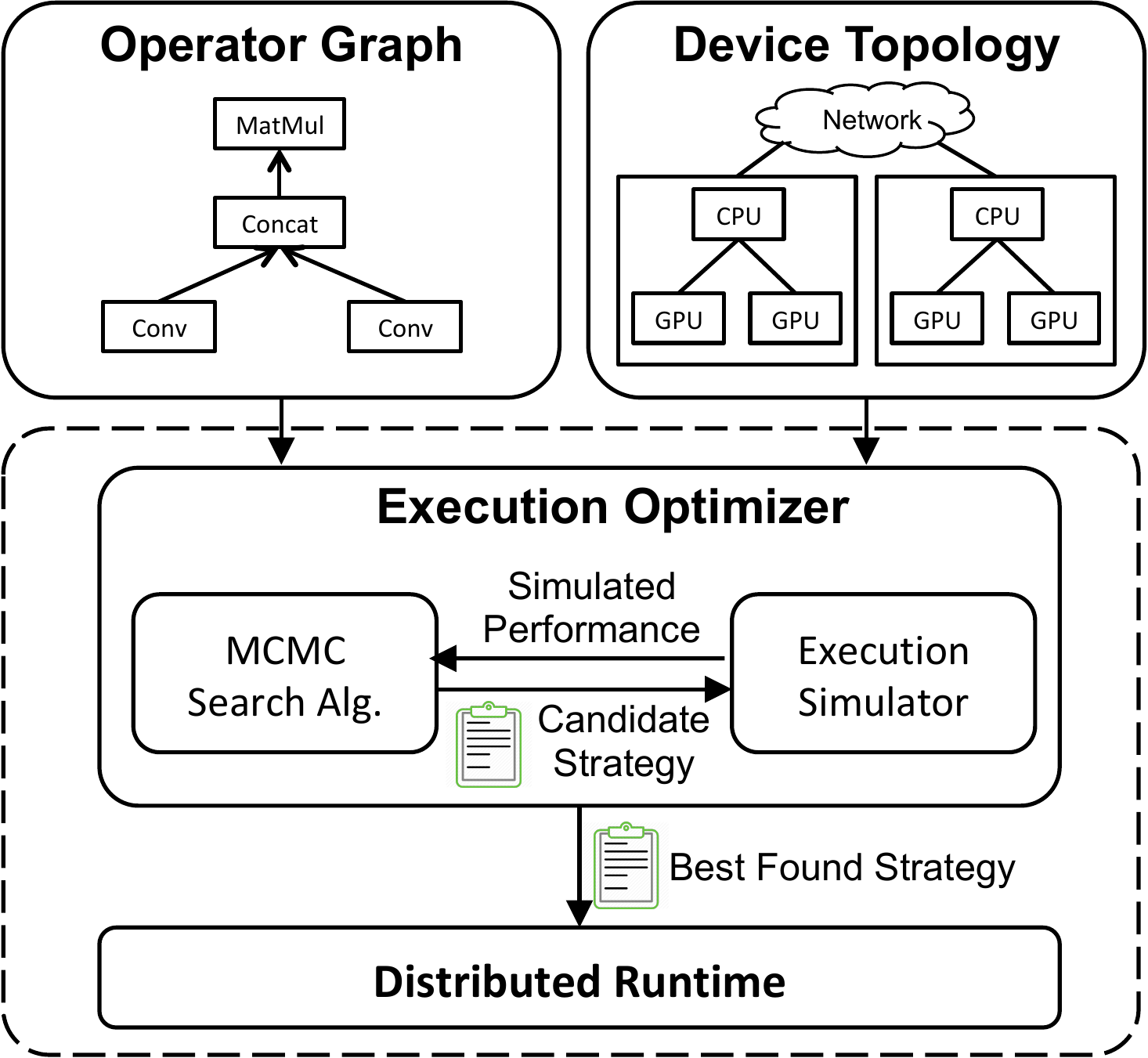}
\vspace{-3mm}
\caption{\Sys overview.}
\label{fig:overview}
\vspace{-4mm}
\end{figure}

The main components of \Sys are shown in Figure~\ref{fig:overview}.
The \Sys {\em execution optimizer} takes an operator graph and a device topology as inputs and automatically generates an efficient parallelization strategy.
The optimizer uses a {\em MCMC search algorithm} to explore the space of possible parallelization strategies and iteratively proposes candidate strategies that are evaluated by a {\em execution simulator}.
The execution simulator uses a {\em delta simulation algorithm} that simulates a new strategy using incremental updates to previous simulations.
The simulated execution time guides the search in generating future candidates.
When the search time budget is exhausted, the execution optimizer sends the best discovered strategy to a {\em distributed runtime} for parallelizing the actual executions.

\subsection{Limitations}
\label{subsec:limitations}
The main limitation of our approach is that the execution simulator assumes the execution time of each operation is predictable and independent of the contents of input tensors, as we discuss in Section~\ref{sec:simulator}.
Therefore, our approach may not be applicable to applications whose execution time is data dependent. 
However, for the DNN applications that are the subject of study here, which are based on dense matrix operations, execution time is highly predictable and independent of the contents of the matrices.

\section{The SOAP Search Space}
\label{sec:problem}
This section introduces the SOAP search space of parallelization strategies for DNNs.
To parallelize a DNN operation across devices, we require each device to compute a disjoint subset of the operation's output tensors. Therefore, we model the parallelization of an operation $o_i$ by
defining how the output tensor of $o_i$ is partitioned.

\begin{table}
\vspace{-3mm}
\caption{Parallelizable dimensions for different operations. The {\em sample} and {\em channel} dimension index different samples and neurons in a tensor, respectively. For 1D and 2D images, the {\em length} and the combination of {\em height} and {\em width} dimensions specify a position in an image.}
\vspace{-3mm}
\label{tab:dimensions}
\centering
\resizebox{\columnwidth}{!}{
\begin{tabular}{|l|lll|}
\hline
\multirow{2}{*}{\bf Operation} & \multicolumn{3}{c|}{\bf Parallelizable Dimensions} \\
& {\bf (S)ample} & {\bf (A)ttribute} & {\bf (P)arameter}\\
\hline
1D pooling & sample & length, channel & \\
1D convolution & sample & length & channel \\
2D convolution & sample & height, width & channel\\
Matrix multiplication & sample & & channel\\
\hline
\end{tabular}
}
\end{table}

\begin{figure}
\centering
\includegraphics[scale=0.145]{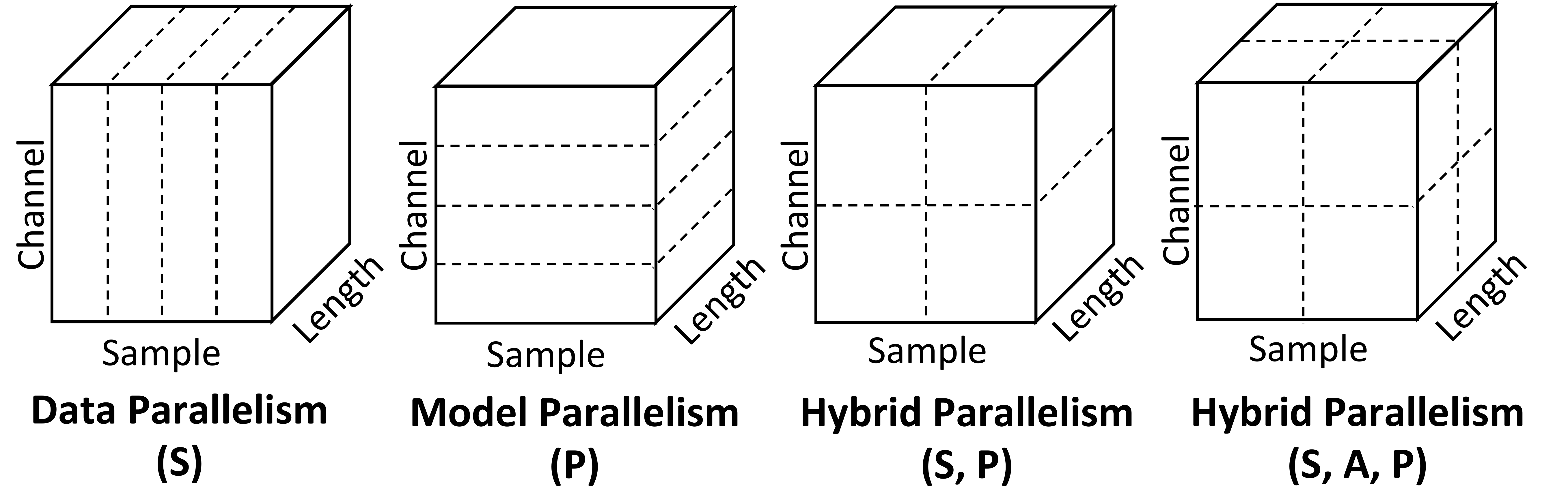}
\vspace{-6mm}
\caption{Example parallelization configurations for 1D convolution. Dashed lines show partitioning the tensor.}
\vspace{-3mm}
\label{fig:example_parallelization}
\end{figure}

For an operation $o_i$, we define its {\em parallelizable dimensions} $\mathcal{P}_i$ as the set of all divisible dimensions in its output tensor. 
$\mathcal{P}_i$ always includes a {\em sample} dimension.
For all other dimensions in $\mathcal{P}_i$, we call it a {\em parameter} dimension if partitioning over that dimension requires splitting the model parameters and call it an {\em attribute} dimension otherwise.
Table~\ref{tab:dimensions} shows the parallelizable dimensions of some example operations. Finally, we also consider parallelism across differ operations in the {\em operation} dimension.


A {\em parallelization configuration} $c_i$ of an operation $o_i$ defines how the operation is parallelized across multiple devices.
Figure~\ref{fig:example_parallelization} shows some example configurations for parallelizing a 1D convolution operation in a single dimension as well as combinations of multiple dimensions. 

For each parallelizable dimension in $\mathcal{P}_i$, $c_i$ includes a positive integer that is the degree of parallelism in that dimension.
$|c_i|$ is the product of the parallelism degrees for all parallelizable dimensions of $c_i$.
We use equal size partitions in each dimension to guarantee well-balanced workload distributions.
A parallelization configuration $c_i$ partitions the operation $o_i$ into $|c_i|$ independent {\em tasks}, denoted as $t_{i:1},...,t_{i:|c_i|}$,
meanwhile $c_i$ also includes the device assignment for each task $t_{i:k}$ ($1\leq k \leq |c_i|$).
Given the output tensor of a task and its operation type, we can infer the necessary input tensors to execute each task. 

\begin{figure}[t]
\vspace{-3mm}
\centering
\includegraphics[scale=0.32]{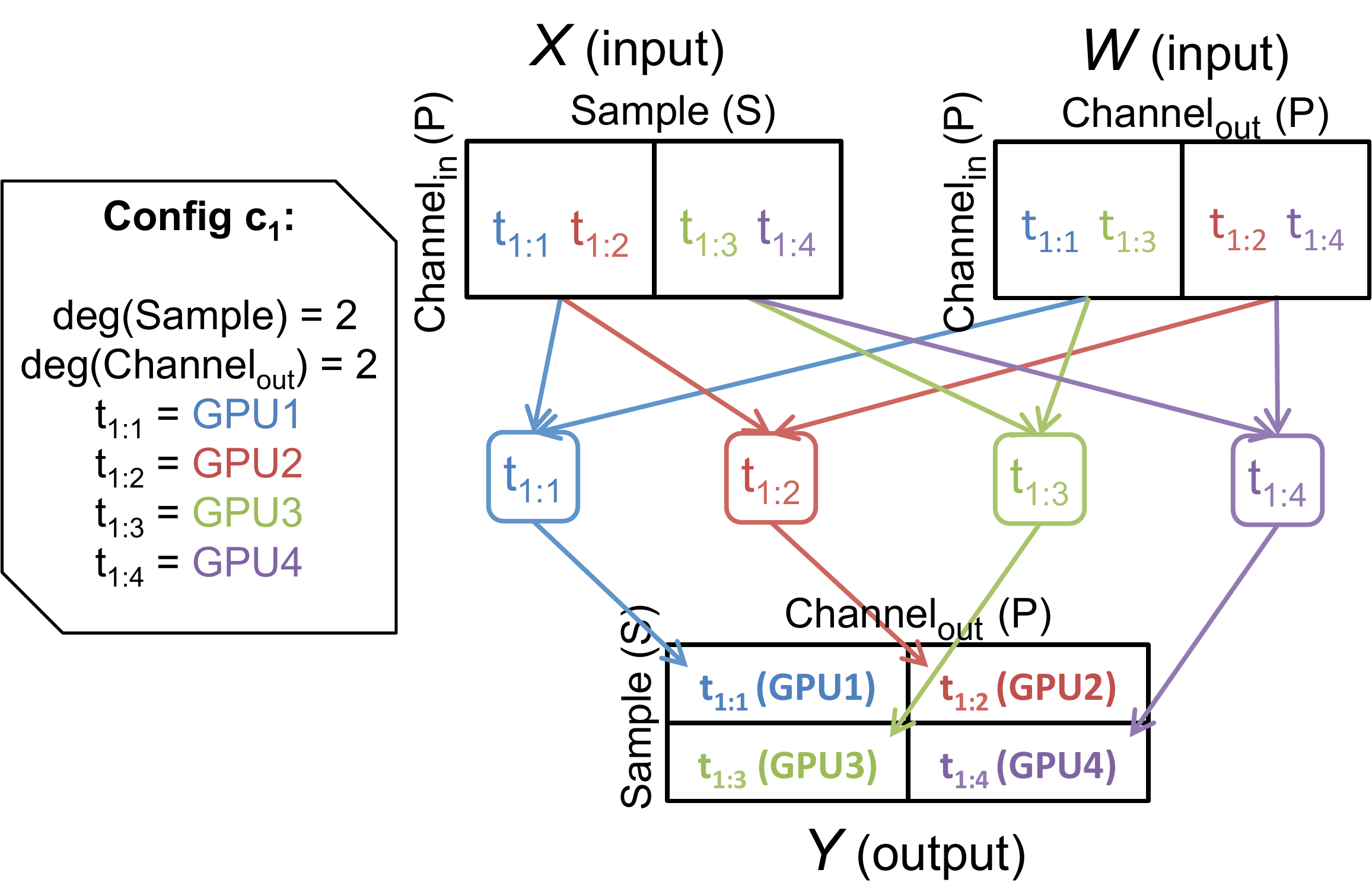}
\vspace{-3mm}
\caption{An example parallelization configuration for a matrix multiplication operation.}
\label{fig:example_configuration}
\vspace{-3mm}
\end{figure}
Figure~\ref{fig:example_configuration} shows an example parallelization configuration for a matrix multiplication operation (i.e., $Y=WX$).
The operation is partitioned into four independent tasks assigned to dedicated GPU devices. The input and output tensors of the tasks are shown in the figure.

A {\em parallelization strategy} $\mathcal{S}$ describes one possible parallelization of an application. 
$\mathcal{S}$ includes a parallelization configuration $c_i$ for each operation $o_i$, and each $o_i$'s configuration can be chosen independently from among all possible configurations for $o_i$.

\section{Execution Simulator}
\label{sec:simulator}
In this section, we describe the execution simulator, which takes an operator graph $\mathcal{G}$, a device topology $\mathcal{D}$, and a parallelization strategy $\mathcal{S}$ as inputs and predicts the execution time to run $\mathcal{G}$ on $\mathcal{D}$ using strategy $\mathcal{S}$.
\Sys simulates the execution process instead of measuring the elapsed time from real executions for two reasons. 
First, processing one iteration of a DNN application can take seconds even on modern GPUs~\cite{LargeSGD, tensorflow}.
The simulator runs up to three orders of magnitude faster than real executions and allows the execution optimizer to explore many more candidates in a given time budget.
Second, the execution simulator requires fewer computation resources. A large-scale execution on thousands of devices can be simulated on a single node. 

The simulator depends on the following assumptions:
\begin{itemize}
\vspace{-1.5mm}
\item[\bf{A1.}] The execution time of each task is predictable with low variance and is independent of the contents of input tensors.
\vspace{-1.5mm}
\item[\bf{A2.}] For each connection $(d_i, d_j)$ between device $d_i$ and $d_j$ with bandwidth $b$, transferring a tensor of size $s$ from $d_i$ to $d_j$ takes $s/b$ time (i.e., the communication bandwidth can be fully-utilized).
\vspace{-1.5mm}
\item[\bf{A3.}] Each device processes the assigned tasks with a FIFO (first-in-first-out) scheduling policy. This is the policy used by modern devices such as GPUs.
\vspace{-1.5mm}
\item[\bf{A4.}] The runtime has negligible overhead. A device begins processing a task as soon as its input tensors are available and the device has finished previous tasks.
\end{itemize}

To simulate an execution, the simulator first builds a {\em task graph}, which includes all tasks derived from operations and dependencies between tasks, and then runs a simulation algorithm to generate an execution timeline.
Section~\ref{subsec:execution_graph} describes task graph construction.
Section~\ref{subsec:base_algorithm} introduces a full simulation algorithm that builds timelines from scratch.
Finally, Section~\ref{subsec:opt_algorithm} introduces an alternative delta simulation algorithm that generates a new timeline using incremental updates to a previous one.

\subsection{Task Graph}
\label{subsec:execution_graph}
\begin{figure*}[t]
\vspace{-6mm}
\subfloat[An example parallelization strategy.] {
\label{fig:global_configuration}
\includegraphics[scale=0.24]{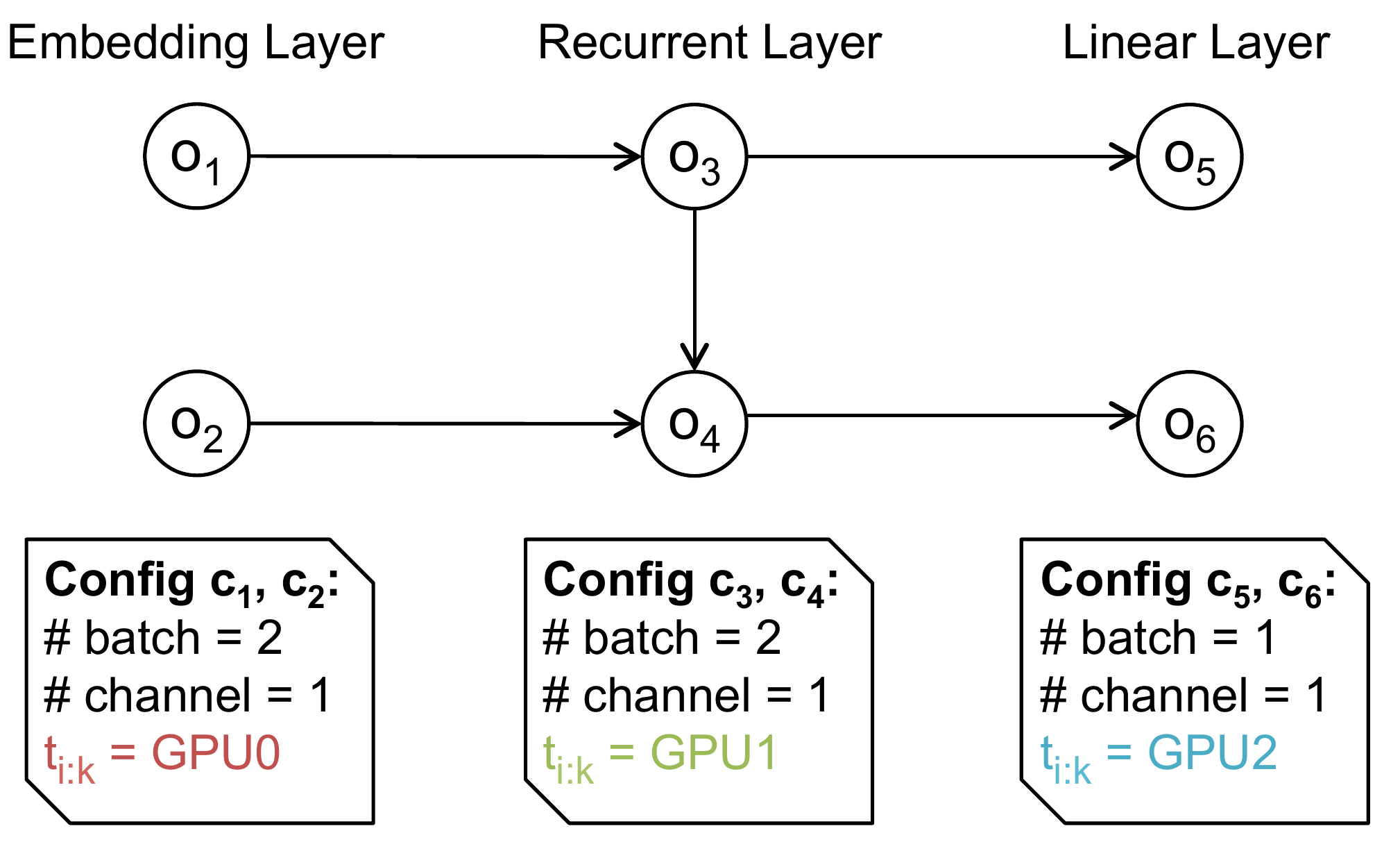}
}
\subfloat[The corresponding task graph.] {
\label{fig:execution_graph}
\includegraphics[scale=0.26]{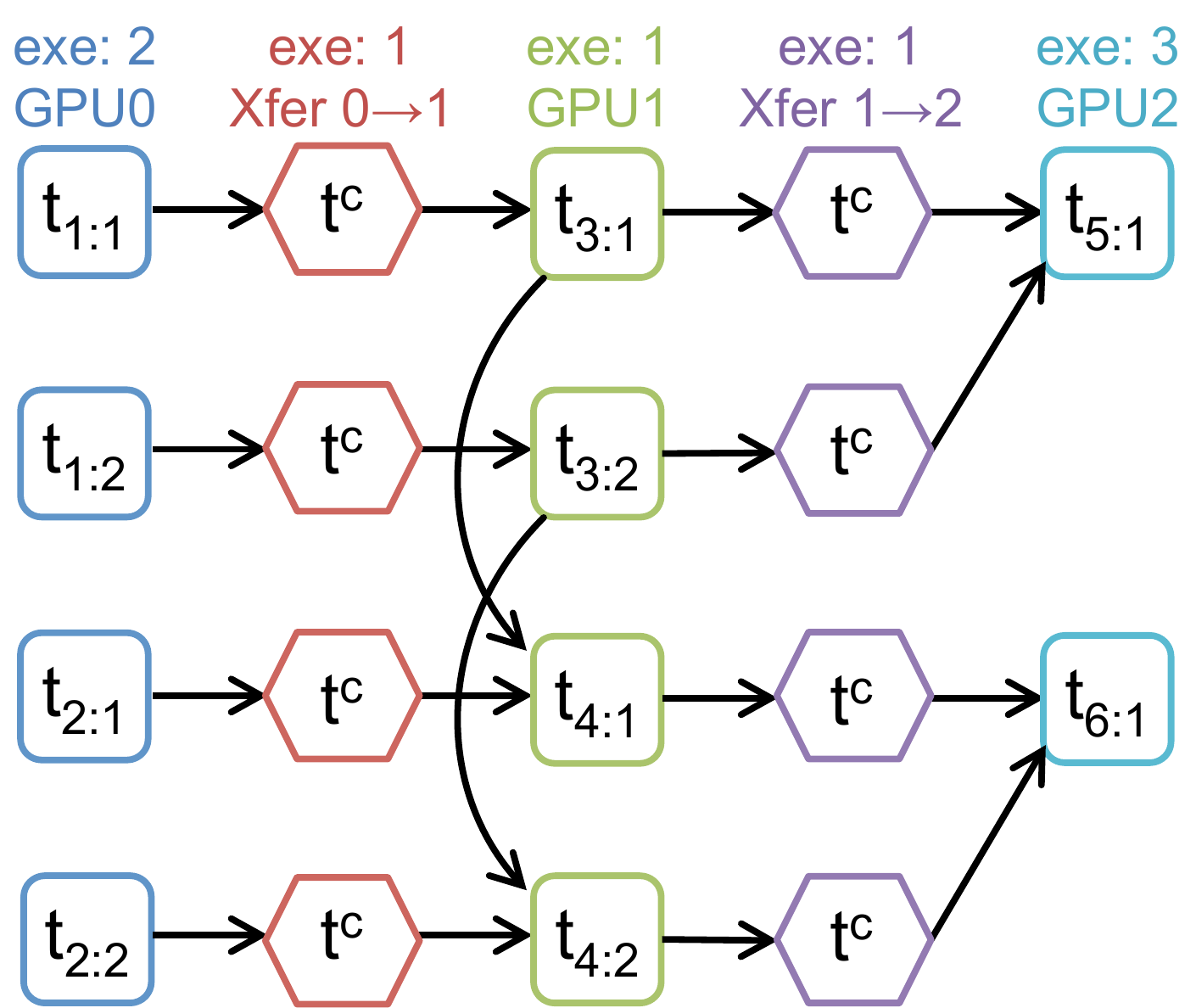}
}
\subfloat[The task graph after the \newline full simulation algorithm.] {
\label{fig:example_simulation_base}
\includegraphics[scale=0.26]{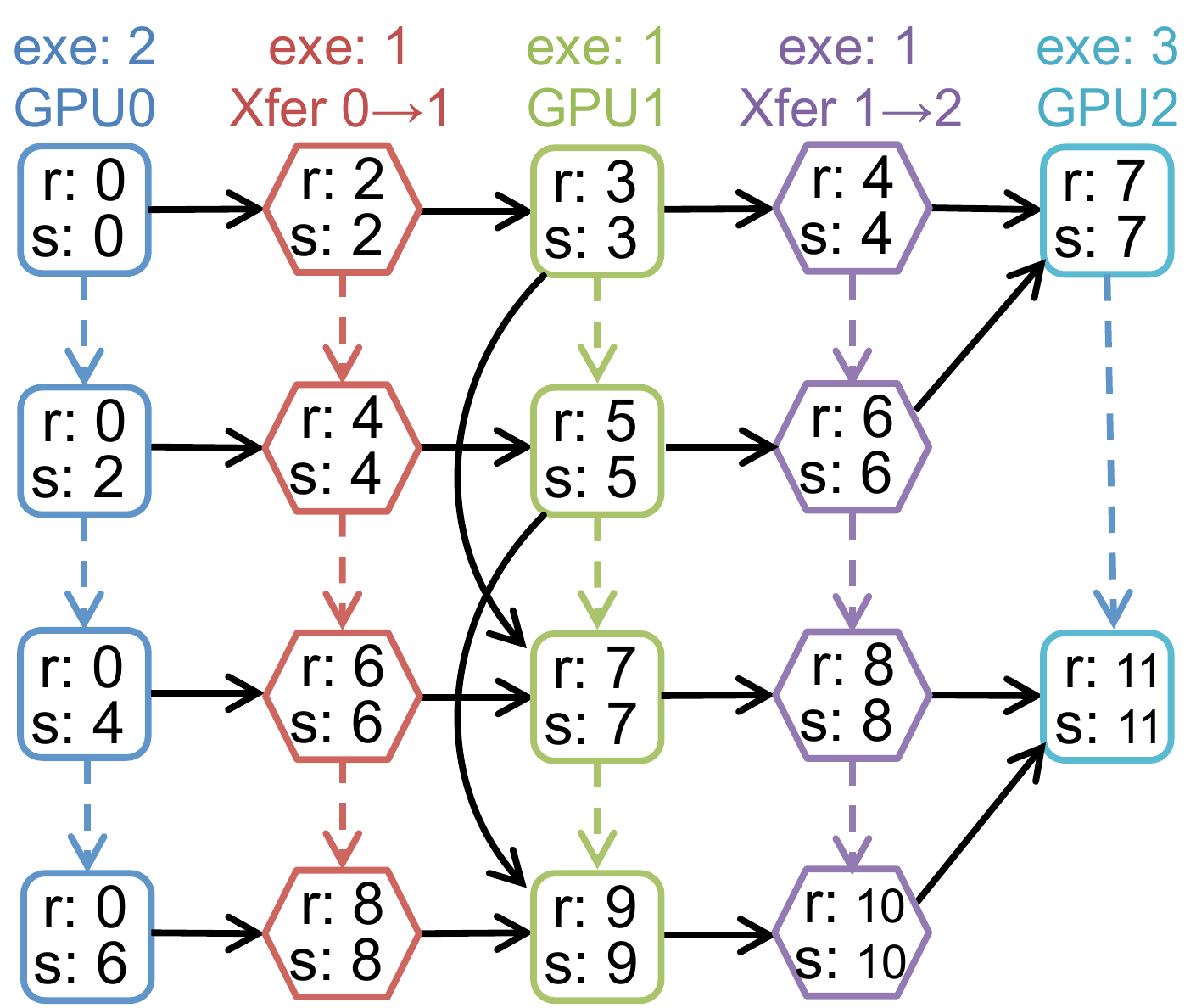}
}
\subfloat[The task graph after the \newline delta simulation algorithm.] {
\label{fig:example_simulation_opt}
\includegraphics[scale=0.26]{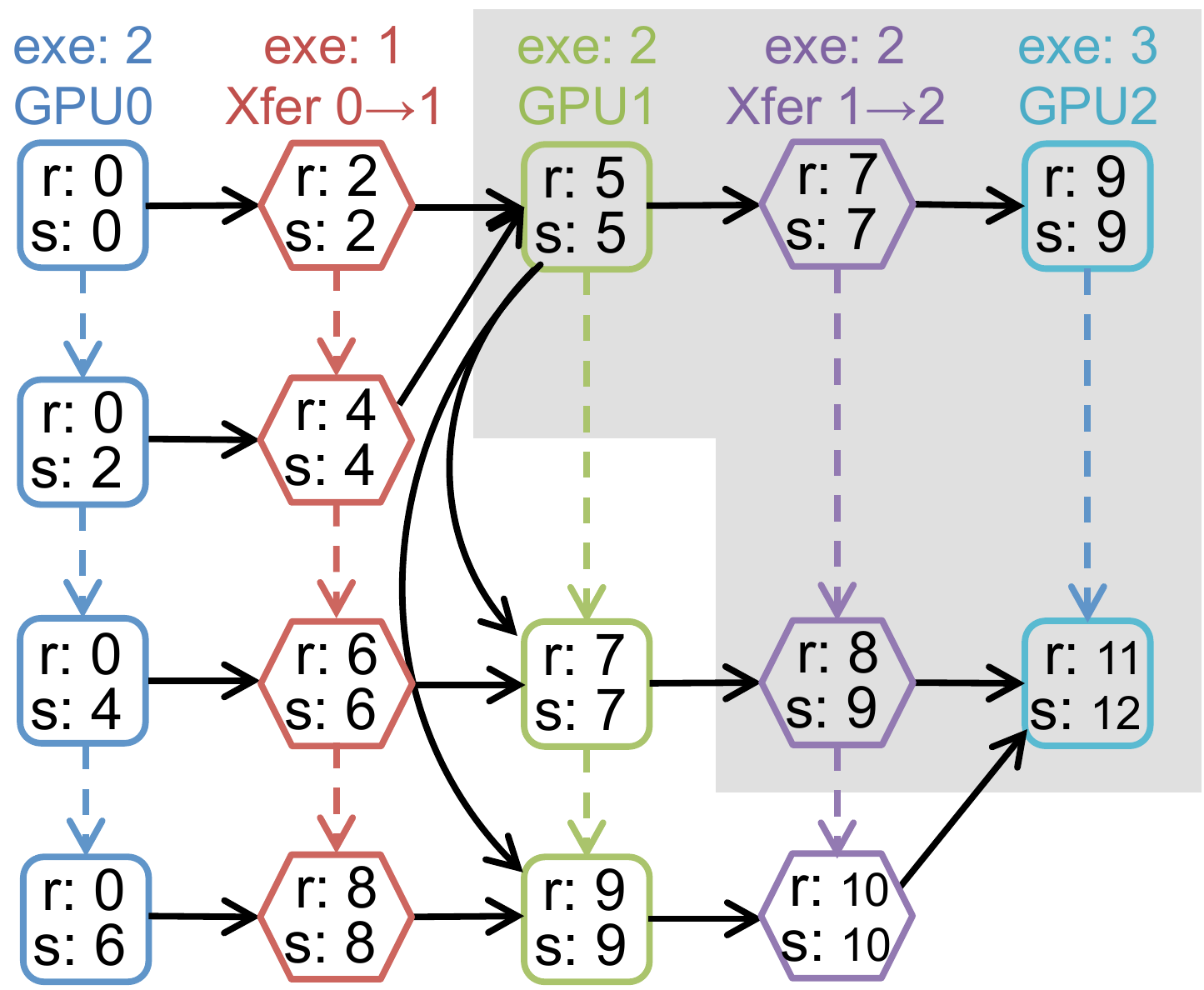}
}
\vspace{-3mm}
\caption{Simulating an example parallelization strategy. The tasks' {\tt exeTime} and {\tt device} are shown on the top of each column.
In Figure~\ref{fig:example_simulation_base} and~\ref{fig:example_simulation_opt}, the word ``r'' and ``s'' indicate the {\tt readyTime} and {\tt startTime} of each task, respectively, and the dashed edges represents the {\tt nextTask}. 
}
\label{fig:example}
\vspace{-3mm}
\end{figure*}

A {\em task graph} models dependencies between individual {\em tasks} derived from operations and can also represent task execution timelines on individual devices.
To unify the abstraction, we treat each hardware connection between devices as a {\em communication device}, and each data transfer as a {\em communication task}.
Note that devices and hardware connections are modeled as separate devices.
This allows computation (i.e., normal tasks) and communication (i.e., communication tasks) to be overlapped if possible.

Given an operator graph $\mathcal{G}$, a device topology $\mathcal{D}$, and a parallelization strategy $\mathcal{S}$, we use the following steps to construct a task graph $\mathcal{T} = (\mathcal{T}_{N}, \mathcal{T}_E)$,
where each node $t\in \mathcal{T}_N$ is a task (i.e., a normal task or a communication task) and each edge $(t_i, t_j) \in \mathcal{T}_E$ is a dependency that task $t_j$ cannot start until task $t_i$ is completed.
Note that the edges in the task graph are simply ordering constraints---the edges do not indicate data flow, as all data flow is included in the task graph as communication tasks.
\begin{itemize}
\vspace{-2mm}
\item[1.] For each operation $o_i \in \mathcal{G}$ with parallelization configuration $c_i$, we add tasks $t_{i:1},...,t_{i:|c_i|}$ into $\mathcal{T}_N$.
\vspace{-2mm}
\item[2.] For each tensor $(o_i, o_j) \in \mathcal{G}$, which is an output of operation $o_i$ and an input of $o_j$, we compute the output sub-tensors written by tasks $t_{i:k_i}$ ($1\leq k_i \leq |c_i|$) and the input sub-tensors read by tasks $t_{j:k_j}$ ($1\leq k_j \leq |c_j|$). 
For every task pair $t_{i:k_i}$ and $t_{j:k_j}$ with shared tensors, if two tasks are assigned to the same device, we add an edge $(t_{i:k_i}, t_{j:k_j})$ into $\mathcal{T}_E$, indicating a dependency between the two tasks, and no communication task is needed. If $t_{i:k_i}$ and $t_{j:k_j}$ with shared tensors are assigned to different devices, we add a communication task $t^c$ to $\mathcal{T}_N$ and two edges $(t_{i:k_i}, t^c)$ and $(t^c, t_{j:k_j})$ to $\mathcal{T}_E$.
The new task $t^c$ is assigned to the communication device between the devices that perform $t_{i:k_i}$ and $t_{j:k_j}$.
\vspace{-2mm}
\end{itemize}

\begin{table}[t]
\vspace{-1mm}
\caption{Properties for each task in the task graph.}
\vspace{-3mm}
\label{tab:properties}
\centering
\resizebox{0.9\columnwidth}{!}{
\begin{tabular}{|l|l|}
\hline
{\bf Property} & {\bf Description} \\
\hline
\multicolumn{2}{|c|}{Properties set in graph construction}\\
\hline
{\tt exeTime} & The elapsed time to execute the task.\\
{\tt device} & The assigned device of the task.\\
$\mathcal{I}$(t) & $\{t_{\er{in}} | (t_{\er{in}}, t) \in \mathcal{T}_E\}$ \\
$\mathcal{O}$(t) & $\{t_{\er{out}} | (t, t_{\er{out}}) \in \mathcal{T}_E\}$ \\
\hline
\multicolumn{2}{|c|}{Properties set in simulation}\\
\hline
{\tt readyTime} & The time when the task is ready to run.\\
{\tt startTime} & The time when the task starts to run.\\
{\tt endTime} & The time when the task is completed.\\
{\tt preTask} & The previous task performed on {\tt device}.\\
{\tt nextTask} & The next task performed on {\tt device}.\\
\hline
\multicolumn{2}{|c|}{Internal properties used by the full simulation algorithm}\\
\hline
\multirow{2}{*}{\tt state} & Current state of the task, which is one of \\
 & {\tt NOTREADY}, {\tt READY}, and {\tt COMPLETE}.\\
\hline
\end{tabular}
}
\vspace{-3mm}
\end{table}

Figure~\ref{fig:global_configuration} shows an example parallelization strategy for a standard 3-layer recurrent neural network consisting of an embedding layer, a recurrent layer, and a linear layer. 
The parallelization strategy represents commonly used model parallelism that assigns operations in each layer to a dedicated GPU.
Figure~\ref{fig:execution_graph} shows the corresponding task graph. Each square and hexagon indicate a normal task and a communication task, respectively, and each directed edge represents a dependency between tasks.

Table~\ref{tab:properties} lists the properties for each task in the task graph.
The {\tt exeTime} property is set during the graph construction.
For a normal task derived from an operation, its {\tt exeTime} is the time to execute the task on the given device and is estimated by running the task multiple times on the device and measuring the average execution time (assumption A1).
A task's {\tt exeTime} is cached, and all future tasks with the same operation type and output size will use the cached value without rerunning the task.
For a communication task, its {\tt exeTime} is the time to transfer a tensor (of size $s$) between devices with bandwidth $b$ and is estimated as $s/b$ (assumption A2).

In addition to the {\tt exeTime} property, \Sys also sets the {\tt device}, $\mathcal{I}(t)$, and $\mathcal{O}(t)$ (defined in Table~\ref{tab:properties}) during graph construction.
Other properties in Table~\ref{tab:properties} remain unset and must be filled in by the simulation. 


\subsection{Full Simulation Algorithm}
\label{subsec:base_algorithm}
\begin{algorithm}[t]
\caption{Full Simulation Algorithm.}
\footnotesize
\begin{algorithmic}[1]
\State {\bf Input}: An operator graph $\mathcal{G}$, a device topology $\mathcal{D}$, and a parallelization strategy $\mathcal{S}$.
\State $\mathcal{T}$ = \Call{BuildTaskGraph}{$\mathcal{G}$, $\mathcal{D}$, $\mathcal{S}$}
\State readyQueue = \{\} {\em // a priority queue sorted by readyTime}
\For {t $\in \mathcal{T}_N$ }
\State t.state = {\tt NOTREADY}
\If {$\mathcal{I}({\rm t}) = \{\}$}
\State t.state = {\tt READY}
\State readyQueue.enqueue(t)
\EndIf
\EndFor
\While{readyQueue $\neq \{\}$}
\State {\bf Task} t = readyQueue.dequeue()
\State {\bf Device} d = t.device
\State t.state = {\tt COMPLETE}
\State t.startTime = $\max$\{t.readyTime, d.last.endTime\}
\State t.endTime = t.startTime + t.exeTime
\State d.last = t
\For {n $\in \mathcal{O}({\rm t})$}
\State n.readyTime = $\max$\{n.readyTime, t.endTime\}
\If {all tasks in $\mathcal{I}({\rm n})$ are {\tt COMPLETE}}
\State n.state = {\tt READY}
\State readyQueue.enqueue(n)
\EndIf
\EndFor
\EndWhile
\State {\bf return} $\max$\{t.endTime $|$ t $\in \mathcal{T}_N$\}
\end{algorithmic}
\label{alg1}
\end{algorithm}

We now describes a full simulation algorithm that we will use as a baseline for comparisons with our delta simulation algorithm. 
Algorithm~\ref{alg1} shows the pseudocode.
It first builds a task graph using the method described in Section~\ref{subsec:execution_graph} and then sets the properties for each task using a variant of Dijkstra's shortest-path algorithm~\cite{IntroAlg}.
Tasks are enqueued into a global priority queue when ready (i.e., all predecessor tasks are completed) and are dequeued in increasing order by their {\tt readyTime}.
Therefore, when a task $t$ is dequeued, all tasks with an earlier {\tt readyTime} have been scheduled, and we can set the properties for task $t$ while maintaining the FIFO scheduling order (assumption A3). 
Figure~\ref{fig:example_simulation_base} shows the execution timeline of the example parallelization strategy.





\subsection{Delta Simulation Algorithm}
\label{subsec:opt_algorithm}
\begin{algorithm}[t]
\caption{Delta Simulation Algorithm.}
\footnotesize
\begin{algorithmic}[1]
\State {\bf Input}: An operator graph $\mathcal{G}$, a device topology $\mathcal{D}$, an original task graph $\mathcal{T}$, and a new configuration $c_i'$ for operation $o_i$.
\State updateQueue = \{\} {\em // a priority queue sorted by readyTime}
\State {\em /*\Call{UpdateTaskGraph}{} returns the updated task graph and a list of tasks with new {\tt readyTime}*/}
\State $\mathcal{T}, \mathcal{L}$ = \Call{UpdateTaskGraph}{$\mathcal{T}$, $\mathcal{G}$, $\mathcal{D}$, $c_i$, $c_i'$}
\State updateQueue.enqueue($\mathcal{L}$)
\While{updateQueue $\neq \{\}$}
\State {\bf Task} t = updateQueue.dequeue()
\State t.startTime = $\max$\{t.readyTime, t.preTask.endTime\}
\State t.endTime = t.startTime + t.exeTime
\For {n $\in \mathcal{O}$(t)}
\If {\Call{UpdateTask}{n}}
\State updateQueue.push(n)
\EndIf
\EndFor
\If {\Call{UpdateTask}{t.nextTask}}
\State updateQueue.push(t.nextTask)
\EndIf
\EndWhile
\State {\bf return} $\max$\{t.endTime $|$ t $\in \mathcal{T}_N$\}
\State
\Function{UpdateTask}{t}
\State t.readyTime = $\max$\{p.endTime $|$ p $\in \mathcal{I}$(t)\}
\State {\em /*Swap t with other tasks on the device to maintain FIFO.*/}
\State t.startTime = $\max$\{t.readyTime, t.preTask.endTime\}
\If {t's readyTime or startTime is changed}
\State {\bf return True}
\Else
\State {\bf return False}
\EndIf
\EndFunction
\end{algorithmic}
\label{alg2}
\end{algorithm}
\Sys uses a MCMC search algorithm that proposes a new parallelization strategy by changing the parallelization configuration of a single operation in the previous strategy (see Section~\ref{subsec:search_algorithm}).
As a result, in the common case, most of the execution timeline does not change from one simulated strategy to the next.
Based on this observation, we introduce a {\em delta simulation algorithm} that starts from a previous task graph and only re-simulates tasks involved in the portion of the execution timeline that changes, an optimization that dramatically speeds up the simulator, especially for strategies for large distributed machines.
The full and delta simulation algorithms always produce the same timeline for a given task graph.

Algorithm~\ref{alg2} shows the pseudocode for the delta simulation algorithm. It first updates tasks and dependencies in the task graph and enqueues all modified tasks into a global priority queue (line 4-5). 
Similar to the Bellman-Ford shortest-path algorithm~\cite{IntroAlg}, the delta simulation algorithm iteratively dequeues updated tasks and propagates the updates to subsequent tasks (line 6-14).



For the example in Figure~\ref{fig:example}, consider a new parallelization strategy derived from the original strategy (Figure~\ref{fig:global_configuration}) by only reducing the parallelism of operation $o_3$ to 1 (i.e., $|c_3|$ = 1).
Figure~\ref{fig:example_simulation_opt} shows the task graph for the new parallelization strategy, which can be generated from the original task graph (in Figure~\ref{fig:example_simulation_base}) by updating the simulation properties of tasks in the grey area.

\section{Execution Optimizer}
\label{sec:optimizer}
\begin{table*}[t]
\vspace{-6mm}
\centering
\caption{Details of the DNNs and datasets used in evaluation.}
\vspace{-3mm}
\label{tab:dnns}
\begin{threeparttable}
\resizebox{\textwidth}{!}{
\begin{tabular}{|l|l||l|l|l|}
\hline
{\bf DNN} & {\bf Description} & {\bf Dataset} & {\bf Reported Acc.} & {\bf Our Acc.} \\
\hline
\multicolumn{5}{|c|}{Convolutional Neural Networks (CNNs)} \\
\hline
AlexNet~\cite{alexnet} & A 12-layer CNN  & Synthetic data & - & - \\
\inception~\cite{inception} & A 102-layer CNN with Inception modules~\cite{GoogleNet} & ImageNet~\cite{imagenet} & 78.0\%\tnote{a} & 78.0\%\tnote{a} \\
\resnet~\cite{resnet} & A 101-layer residual CNN with shortcut connections & ImageNet~\cite{imagenet} & 76.4\%\tnote{a} & 76.5\%\tnote{a}\\
\hline
\multicolumn{5}{|c|}{Recurrent Neural Networks (RNNs)}\\
\hline
RNNTC~\cite{RNNTC} & 4 recurrent layers followed by a softmax layer & Movie Reviews~\cite{moviereviews} & 79.8\% & 80.3\%\\
RNNLM~\cite{RNNLM} & 2 recurrent layers followed by a softmax layer & Penn Treebank~\cite{penntreebank} & 78.4\tnote{b} & 76.1\tnote{b}\\
NMT~\cite{GNMT} & 4 recurrent layers followed by an attention and a softmax layer & WMT English-German~\cite{wmt16} & 19.67\tnote{c} & 19.85\tnote{c} \\
\hline
\end{tabular}
}
\begin{tablenotes}
\footnotesize
\item[a] top-1 accuracy for single crop on the validation dataset (higher is better). 
\item[b] word-level test perplexities on the Peen Treebank dataset (lower is better).
\item[c] BLEU scores~\cite{bleu} on the test dataset (higher is better).
\end{tablenotes}
\end{threeparttable}
\vspace{-5mm}
\end{table*}

This section describes the {\em execution optimizer} that takes an operator graph and a device topology as inputs and automatically finds an efficient parallelization strategy.
Using the simulator as an oracle, \Sys transforms the parallelization optimization problem into a cost minimization problem, namely minimizing the predicted execution time.
The primary advantage of this approach is that it avoids explicitly encoding the trade-offs between interdependent optimizations (e.g., reducing data transfers v.s. balancing workload distributions) and simply focuses on minimizing the application's overall execution time.

Finding the optimal parallelization strategy is NP-hard, by an easy reduction from {\em minimum makespan}~\cite{lam1977worst}.
In addition, as described in Section~\ref{sec:problem}, the number of possible strategies is exponential to the number of operations in the operator graph, which makes it intractable to exhaustively enumerate the search space.
To find a low-cost strategy, \Sys uses a cost minimization search procedure to heuristically explore the space and returns the best strategy discovered. 

\subsection{MCMC Sampling}
This section briefly introduces the MCMC sampling method used by the execution optimizer. 
MCMC sampling is a technique for obtaining samples from a probability distribution so that higher probability samples are visited proportionately more often than low probability samples. 
A common method (described in~\cite{MCMC}) to transform a cost function ${\er cost}(\cdot)$ into a probability distribution is the following, where $\beta$ is a constant that can be chosen:
\begin{equation}
p(\mathcal{S}) \propto \exp\big( -\beta \cdot \er{cost}(\mathcal{S})\big)
\end{equation}

MCMC works by starting at any point in the search space (a random point, or perhaps a well-known starting point) and then generating a sequence of points with the guarantee that in the limit the set of points visited approaches the distribution given by $p(\cdot)$. 
In our setting, we begin with some parallelization strategy $\mathcal{S}_0$ and then generate a sequence of strategies $\mathcal{S}_0, \mathcal{S}_1,\ldots$.

We use the Metropolis-Hastings algorithm~\cite{MetropolisHastings} for generating Markov chains, which maintains a current strategy $\mathcal{S}$ and {\em proposes} a modified strategy $\mathcal{S}^*$ from a proposal distribution $q(\mathcal{S}|\mathcal{S}^*)$.
If the proposal is {\em accepted}, $\mathcal{S}^*$ becomes the new current strategy, otherwise another strategy based on $\mathcal{S}$ is proposed. 
This process is repeated indefinitely (e.g., until a time budget is exhausted).
If the proposal distribution is symmetric, $q(\mathcal{S} | \mathcal{S}^*) = q(\mathcal{S}^* | \mathcal{S})$, the acceptance criteria of a new strategy is the following:
\begin{equation}
\label{eqn2}
\begin{split}
& \alpha(\mathcal{S}\rightarrow \mathcal{S}^*)  = \min\big(1, p(\mathcal{S}^*) / p(\mathcal{S})\big) \\
& =  \min\Big(1, \exp\big(\beta \cdot(\er{cost}(\mathcal{S}) - \er{cost}(\mathcal{S}^*)\big)\Big)
\end{split}
\end{equation}

The acceptance criteria has several important properties. If $\mathcal{S}^*$ has a lower cost than $\mathcal{S}$, then $\mathcal{S}^*$ is always accepted. 
If $\mathcal{S}^*$ has a higher cost than $\mathcal{S}$,  then $\mathcal{S}^*$ may still be accepted with a probability that decreases as a function of the difference between $\er{cost}(\mathcal{S})$ and $\er{cost}(\mathcal{S}^*)$.
Intuitively, MCMC tends to behave as a greedy search algorithm, preferring to move towards lower cost whenever that is readily available, but can also escape local minima. 

\subsection{Search Algorithm}
\label{subsec:search_algorithm}
Our method for generating proposals is simple: an operation in the current parallelization strategy is selected at random, and its parallelization configuration is replaced by a random configuration.
Our definition of the proposal distribution $q(\cdot)$ satisfies the symmetry property, $q(\mathcal{S} | \mathcal{S}^*) = q(\mathcal{S}^* | \mathcal{S})$, since, for any operation, its configurations are selected with the same probability.


We uses existing strategies (e.g., data parallelism, expert-designed strategies) as well as randomly generated strategies as the initial candidates for the search algorithm. 
For each initial strategy, the search algorithm iteratively proposes new candidates until one of the following two criteria is satisfied: (1) the search time budget for current initial strategy is exhausted; or (2) the search procedure cannot further improve the best discovered strategy for half of the search time. 


\section{\Sys Runtime}
\label{sec:impl}
We found that existing deep learning systems (e.g., TensorFlow~\cite{tensorflow}, PyTorch~\cite{pytorch}, Caffe2~\cite{caffe2}, and MXNet~\cite{mxnet}) only support parallelizing an operation in the batch dimension through data parallelism, and it is non-trivial to parallelize an operation in other dimensions or combinations of several dimensions in these systems. 
In addition, we are not aware of any existing system that supports parallelization at the granularity of individual operations.

To support parallelizing DNN models using any strategy defined in our parallelization space (see Section~\ref{sec:problem}), we implemented the \Sys distributed runtime in Legion~\cite{Legion12}, a high-performance parallel runtime for distributed heterogeneous architectures, and use cuDNN~\cite{cudnn} and cuBLAS~\cite{cublas} as the underlying libraries for processing DNN operations. 
We use the Legion high-dimensional partitioning interface~\cite{dependentpartitioning} to support parallelizing an operation in any combination of the parallelizable dimensions and use Legion's fine-grain control mechanism to control parallelization at the granularity of each operation.

The key difference between the \Sys runtime and existing systems is that \Sys supports parallelizing an operation in any combination of the parallelizable dimensions and controls parallelization at the granularity of individual operations.

\section{Evaluation}
\label{sec:eval}
This section evaluates the performance of \Sys on six real-world DNN benchmarks and two GPU clusters.
Section~\ref{subsec:setup} describes the experimental setup for the evaluation. 
Section~\ref{subsec:parallelization_performance} compares \Sys with state-of-the-art parallelization approaches. 
Section~\ref{subsec:eval_simulator} evaluates the accuracy and efficiency of the execution simulator.
Sections~\ref{subsec:eval_optimizer} and~\ref{subsec:case_study} evaluate the quality of the best strategies discovered by the execution optimizer and discuss two of the best discovered strategies. 

\subsection{Experimental Setup}
\label{subsec:setup}

Table~\ref{tab:dnns} summarizes the DNNs used in our experiments.
AlexNet, \inception, and \resnet are three CNNs that achieved the best accuracy in the ILSVRC competitions~\cite{ILSVRC15}. 
For AlexNet, the per-iteration training time is smaller than the time to load training data from disk.
We follow the suggestions in~\cite{tensorflowbenchmark} and use synthetic data to benchmark the performance of AlexNet.
For all other experiments, the training data is loaded from disk in the training procedure.

RNNTC, RNNLM and NMT are sequence-to-sequence RNN models for text classification, language modeling, and neural machine translation, respectively. 
RNNTC uses four LSTM layers with a hidden size of 1024. RNNLM uses two LSTM layers with a hidden size of 2048.
Both RNN models include a softmax linear after the last LSTM layer.
NMT includes an encoder and a decoder, both of which consist of 2 LSTM layers with a hidden size of 1024.
To improve model accuracy, we also use an attention layer~\cite{NMT2} on top of the last decoder LSTM layer. 
Figure~\ref{fig:case_study_nmt} illustrates the structure of the NMT model.
For all three RNN models, we set the number of unrolling steps for each recurrent layer to 40.

We follow prior work~\cite{alexnet, inception, resnet, RNNTC, RNNLM, GNMT} to construct operator graphs and set hyperparameters (e.g., learning rates, weight decays).
We use synchronous training and a batch size of 64 for all DNN benchmarks, except for AlexNet, which uses a batch size of 256.



\begin{figure}[t]
\vspace{-6mm}
\centering
\subfloat[The P100 Cluster (4 nodes).] {
\label{fig:sherlock}
\includegraphics[scale=0.25]{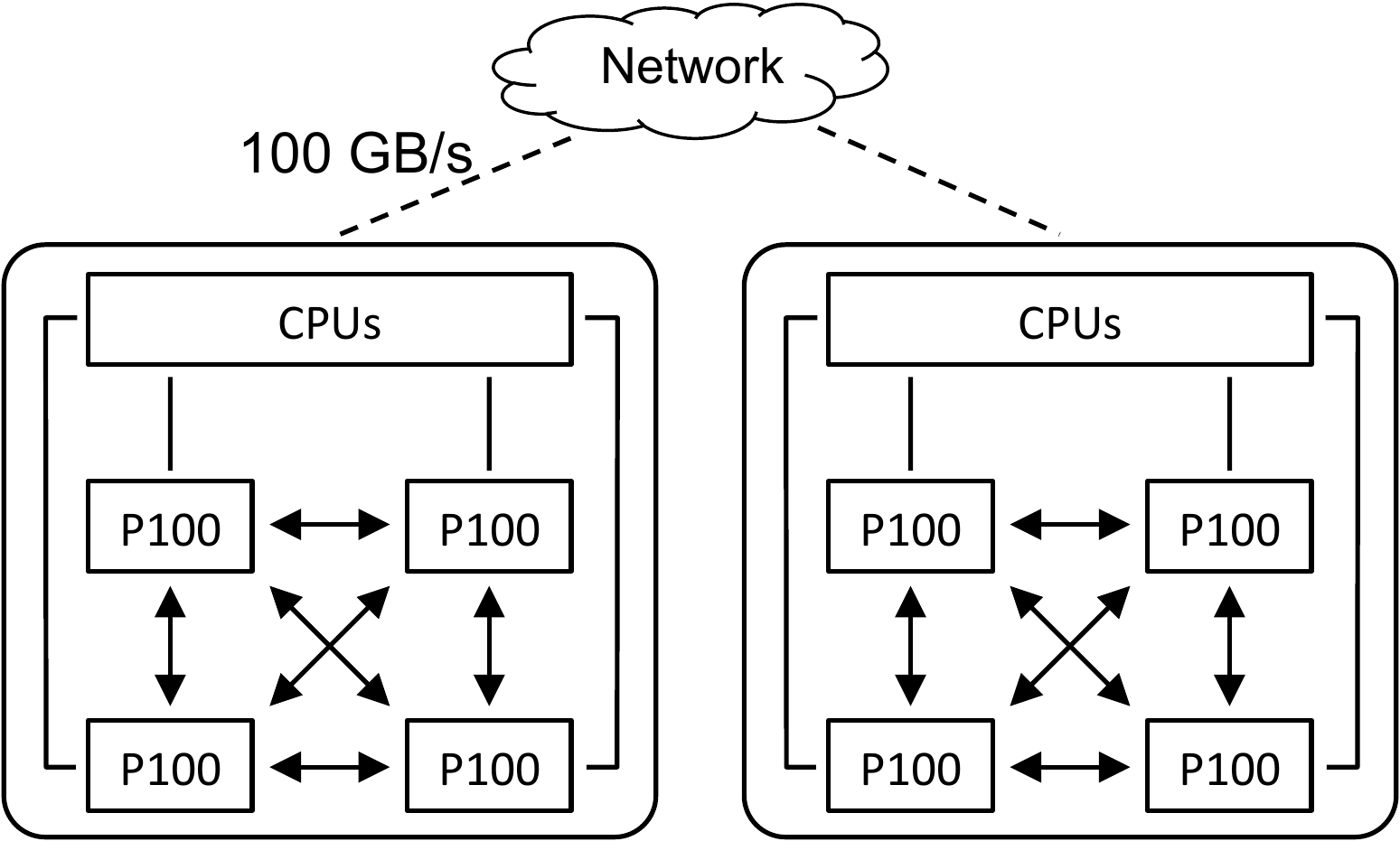}
}
\subfloat[The K80 Cluster (16 nodes).] {
\label{fig:xstream_arch}
\includegraphics[scale=0.25]{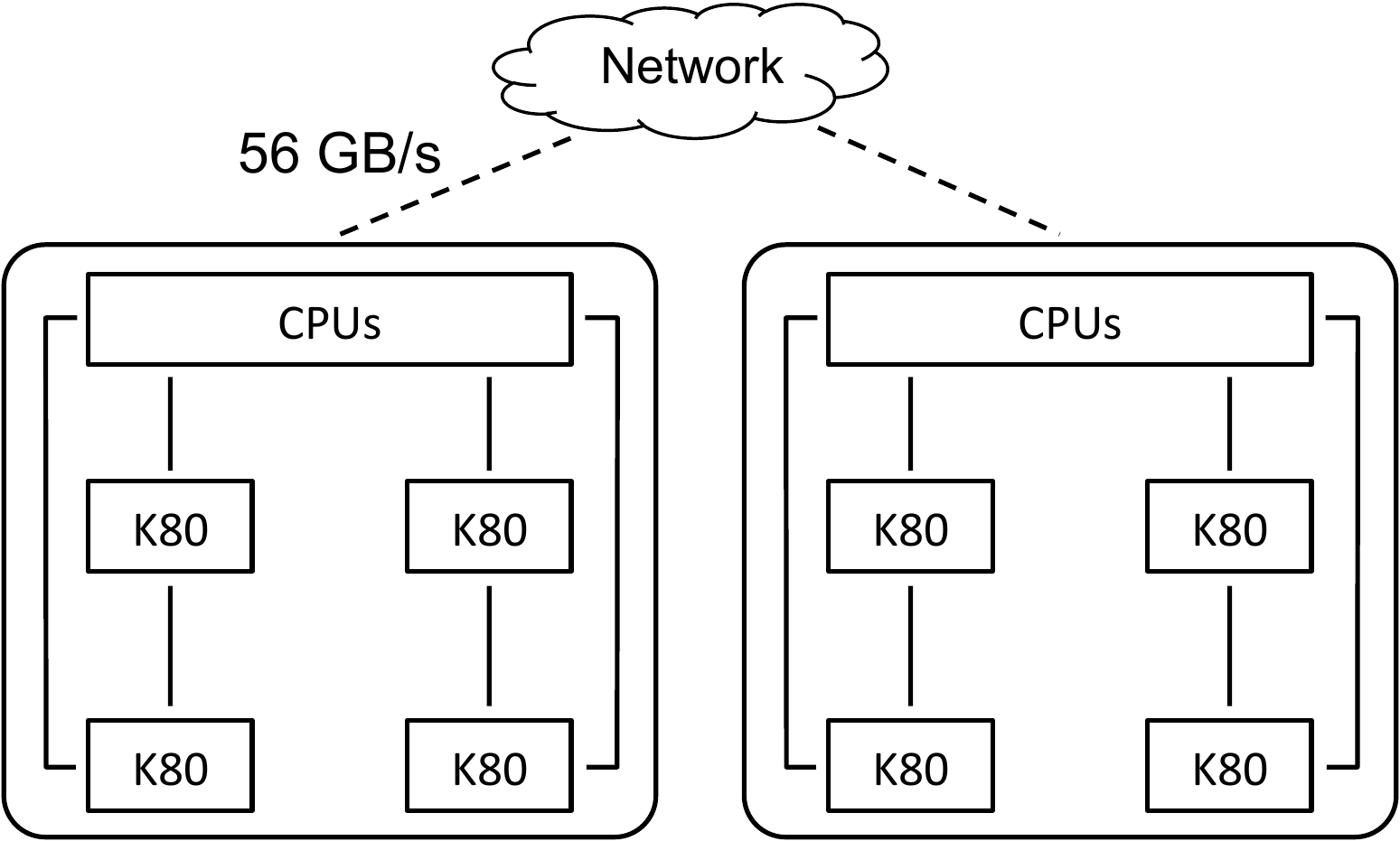}
}
\vspace{-3mm}
\caption{Architectures of the GPU clusters used in the experiments. 
An arrow line indicates a NVLink connection. A solid line is a PCI-e connection. Dashed lines are Infiniband connections across different nodes.}
\label{fig:cluster_overview}
\vspace{-3mm}
\end{figure}
To evaluate the performance of \Sys with different device topologies, we performed the experiments on two GPU clusters, as shown in Figure~\ref{fig:cluster_overview}.
The first cluster contains 4 compute nodes, each of which is equipped with two Intel 10-core E5-2600 CPUs, 256GB main memory, and four NVIDIA Tesla P100 GPUs. GPUs on the same node are connected by NVLink, and nodes are connected over 100GB/s EDR Infiniband.
The second cluster consists of 16 nodes, each of which is equipped with two Intel 10-core E5-2680 GPUs, 256GB main memory, and four NVIDIA Tesla K80 GPUs. Adjacent GPUs are connected by a separate PCI-e switch, and all GPUs are connected to CPUs through a shared PCI-e switch. Compute nodes in the cluster are connected over 56 GB/s EDR Infiniband.

Unless otherwise stated, we set 30 minutes as the time budget for the execution optimizer and use data parallelism and a randomly generated parallelization strategy as the initial candidates for the search algorithm.
As shown in Section~\ref{subsec:simulator_execution_time}, the search procedure terminates in a few minutes for most executions.

\subsection{Parallelization Performance}
\label{subsec:parallelization_performance}
\subsubsection{Per-iteration Performance}
\begin{figure}[t]
\vspace{-5mm}
\centering
\includegraphics[scale=0.35]{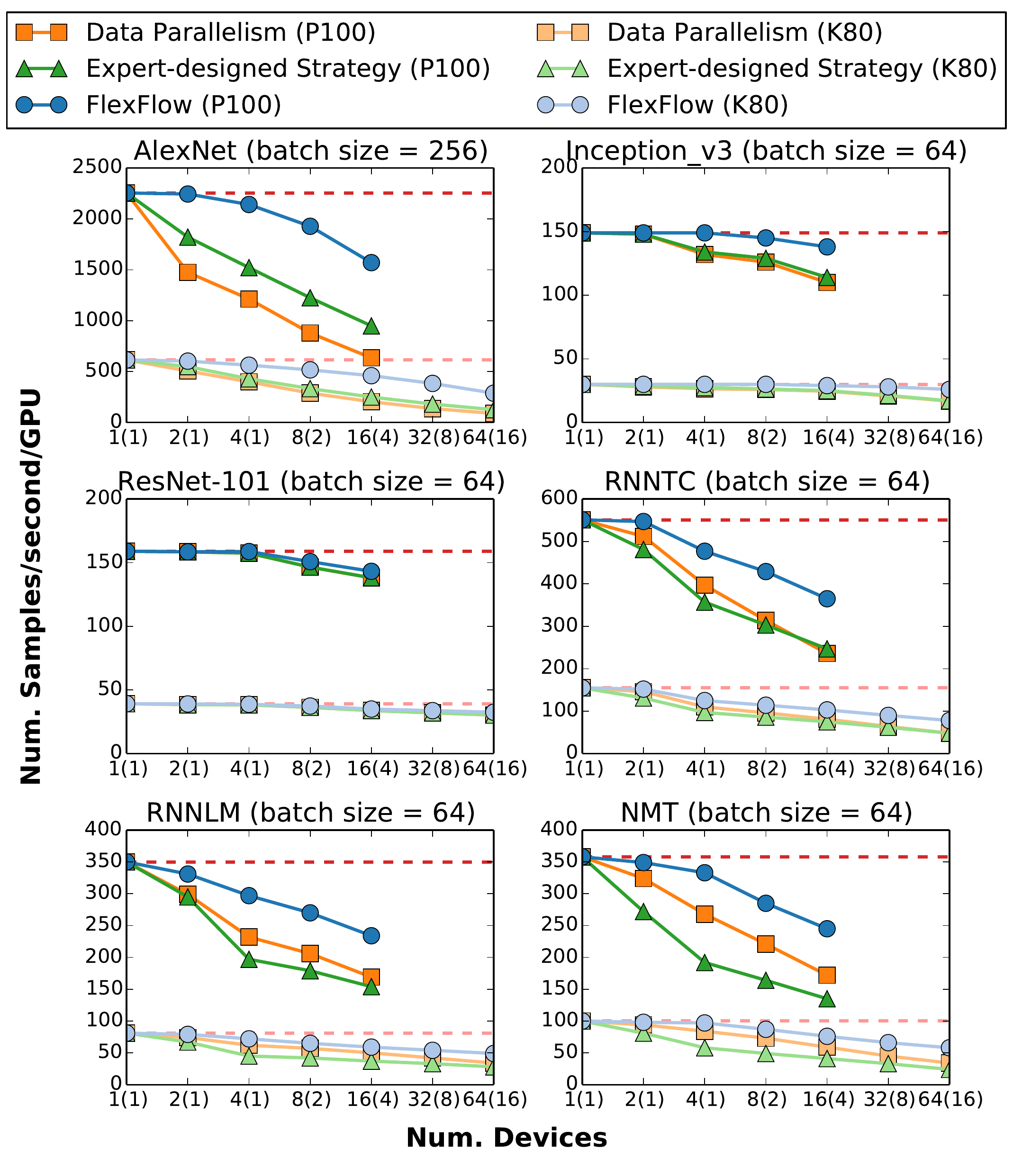}
\vspace{-5mm}
\caption{Per-iteration training performance on six DNN benchmarks. Numbers in parenthesis are the number of compute nodes used in the experiments. The dash lines show the ideal training throughput.}
\label{fig:all_training_throughput}
\vspace{-3mm}
\end{figure}
We compare the per-iteration training performance of \Sys with the following baselines.
Data parallelism is commonly used in existing deep learning systems~\cite{tensorflow, caffe2, pytorch}. 
To control for implementation differences, we ran data parallelism experiments in TensorFlow r1.7, PyTorch v0.3, and our implementation and compared the performance numbers.
Compared to TensorFlow and PyTorch, \Sys achieves the same or better performance numbers on all six DNN benchmarks, and therefore we report the data parallelism performance achieved by \Sys in the experiments. 

Expert-designed strategies optimize parallelization based on domain experts' knowledge and experience.
For CNNs,~\cite{OWT} uses data parallelism for parallelizing convolutional and pooling layers and switches to model parallelism for densely-connected layers.
For RNNs,~\cite{GNMT} uses data parallelism that replicates the entire operator graph on each compute node and uses model parallelism that assign operations with the same depth to the same GPU on each node.
These expert-designed strategies are used as a baseline in our experiments.
Model parallelism only exposes limited parallelism by itself, and we compare against model parallelism as a part of these expert-designed strategies.

Figure~\ref{fig:all_training_throughput} shows the per-iteration training performance on all six DNN benchmarks.
For \resnet, \Sys finds strategies similar to data parallelism (except using model parallelism on a single node for the last fully-connected layer) and therefore achieves similar parallelization performance.
For other DNN benchmarks, \Sys finds more efficient strategies than the baselines and achieves 1.3-3.3$\times$ speedup. 
Note that \Sys performs the same operations as data parallelism and expert-designed strategies, and the performance improvement is achieved by using faster parallelization strategies.
We found that the parallelization strategies discovered by \Sys have two advantages over data parallelism and expert-designed strategies.
\begin{figure}[t]
\vspace{-10mm}
\centering
\subfloat[Per-iteration\newline execution time.] {
\label{fig:nmt_exe_time}
\includegraphics[scale=0.18]{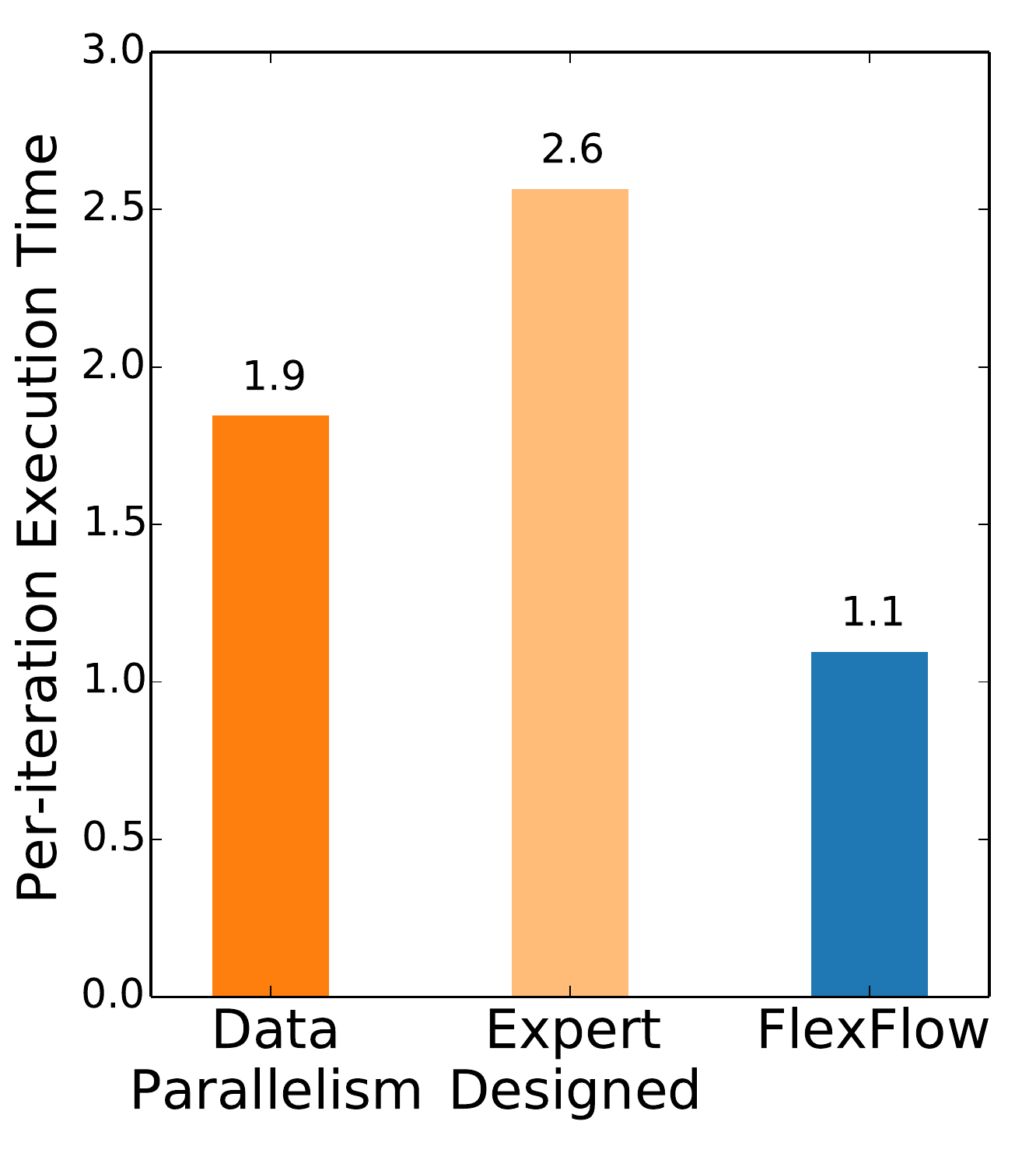}
}
\subfloat[Overall data trans-\newline fers per iteration.] {
\label{fig:nmt_data_xfer}
\includegraphics[scale=0.18]{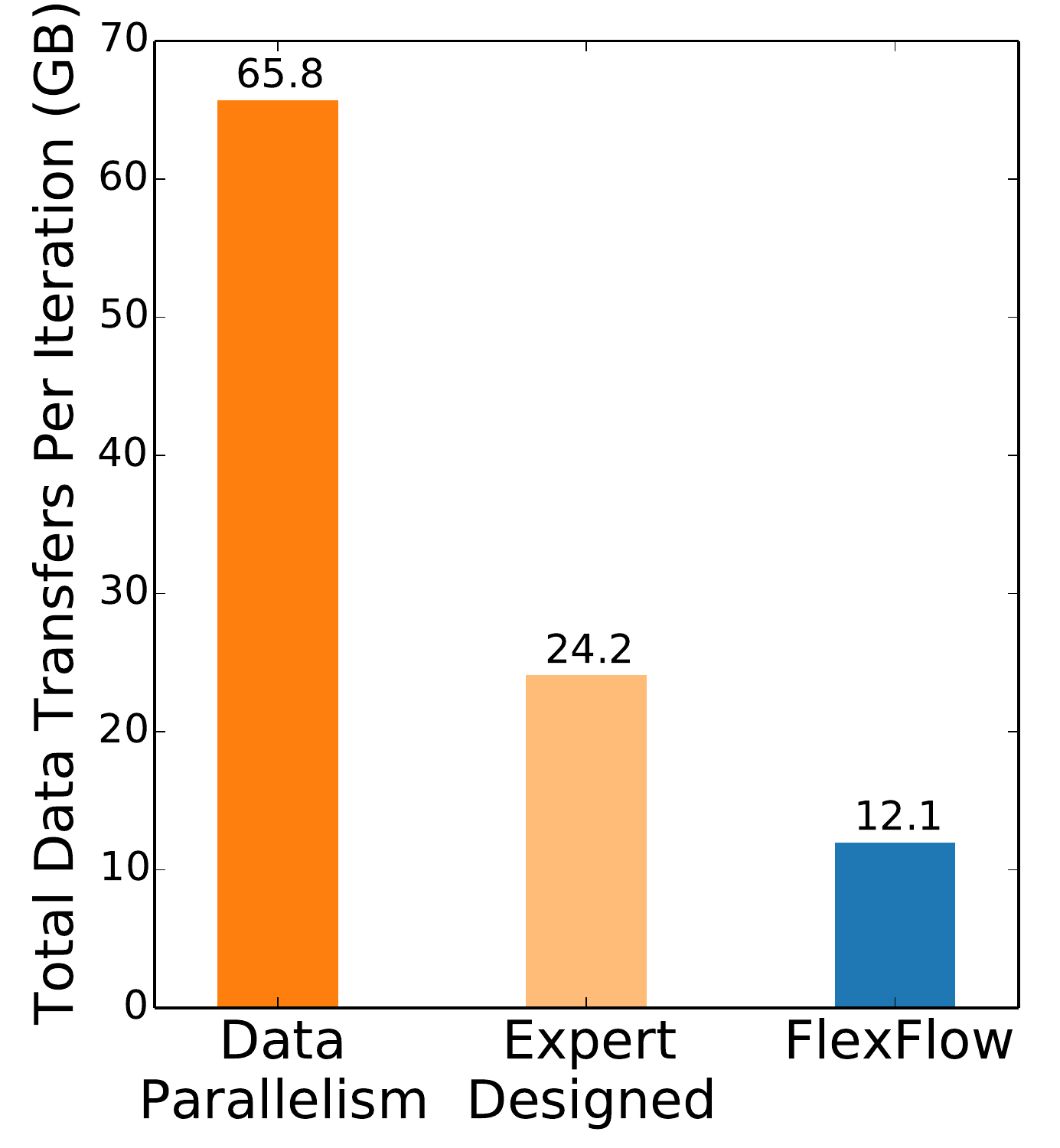}
}
\subfloat[Overall task computation time per iteration.] {
\label{fig:nmt_compute_time}
\includegraphics[scale=0.18]{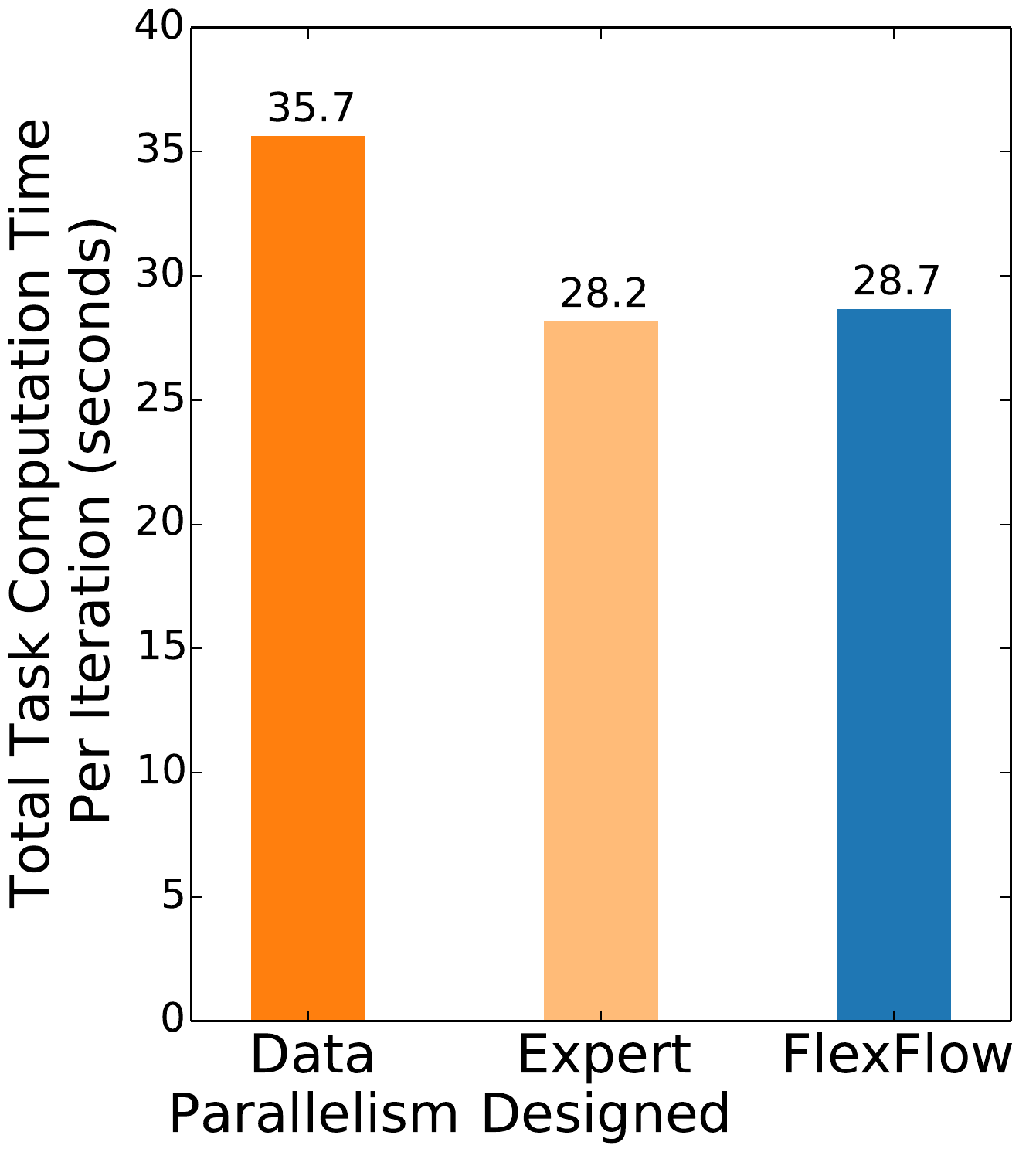}
}
\vspace{-3mm}
\caption{Parallelization performance for the NMT model on 64 K80 GPUs (16 nodes).
\Sys reduces per-iteration execution time by 1.7-2.4$\times$ and data transfers by 2-5.5$\times$ compared to other approaches.
\Sys achieves similar overall task computation time as expert-designed strategy, which is 20\% fewer than data parallelism.
}
\vspace{-3mm}
\end{figure}

{\bf Reducing overall communication costs.} Similar to existing deep learning systems, the \Sys distributed runtime supports overlapping data transfers with computation to hide communication overheads.
However, as we scale the number of devices, the communication overheads increase, but the computation time used to hide communication remains constant. 
Therefore, reducing overall communication costs is beneficial for large-scale distributed training. 
Figure~\ref{fig:nmt_data_xfer} shows that, to parallelize the NMT model on 64 K80 GPUs (16 nodes), \Sys reduces the per-iteration data transfers by 2-5.5$\times$ compared to other parallelization approaches.

{\bf Reducing overall task computation time.} Data parallelism always parallelizes an operation in the batch dimension.
However, as reported in~\cite{OptCNN}, parallelizing an operation through different dimensions can result in different task computation time.
For the matrix multiplication operation in the NMT model, parallelizing it in the channel dimension reduces the operation's overall computation time by 38\% compared to parallelizing the operation in the batch dimension.
Figure~\ref{fig:nmt_compute_time} shows that \Sys reduces the overall task computation time by 20\% compared to data parallelism for the NMT model.
The expert-designed strategy achieves slightly better total task computation time than \Sys. 
However, this is achieved by using model parallelism on each node, which disables any parallelism within each operation and results in imbalanced workloads. 
As a result, the expert-designed strategy achieves even worse execution performance than data parallelism (see Figure~\ref{fig:nmt_exe_time}).
\Sys reduces the overall task computation time while enabling parallelism within an operation and maintaining load balance.


\subsubsection{End-to-end Performance}
\begin{figure}
\vspace{-8mm}
\centering
\includegraphics[scale=0.24]{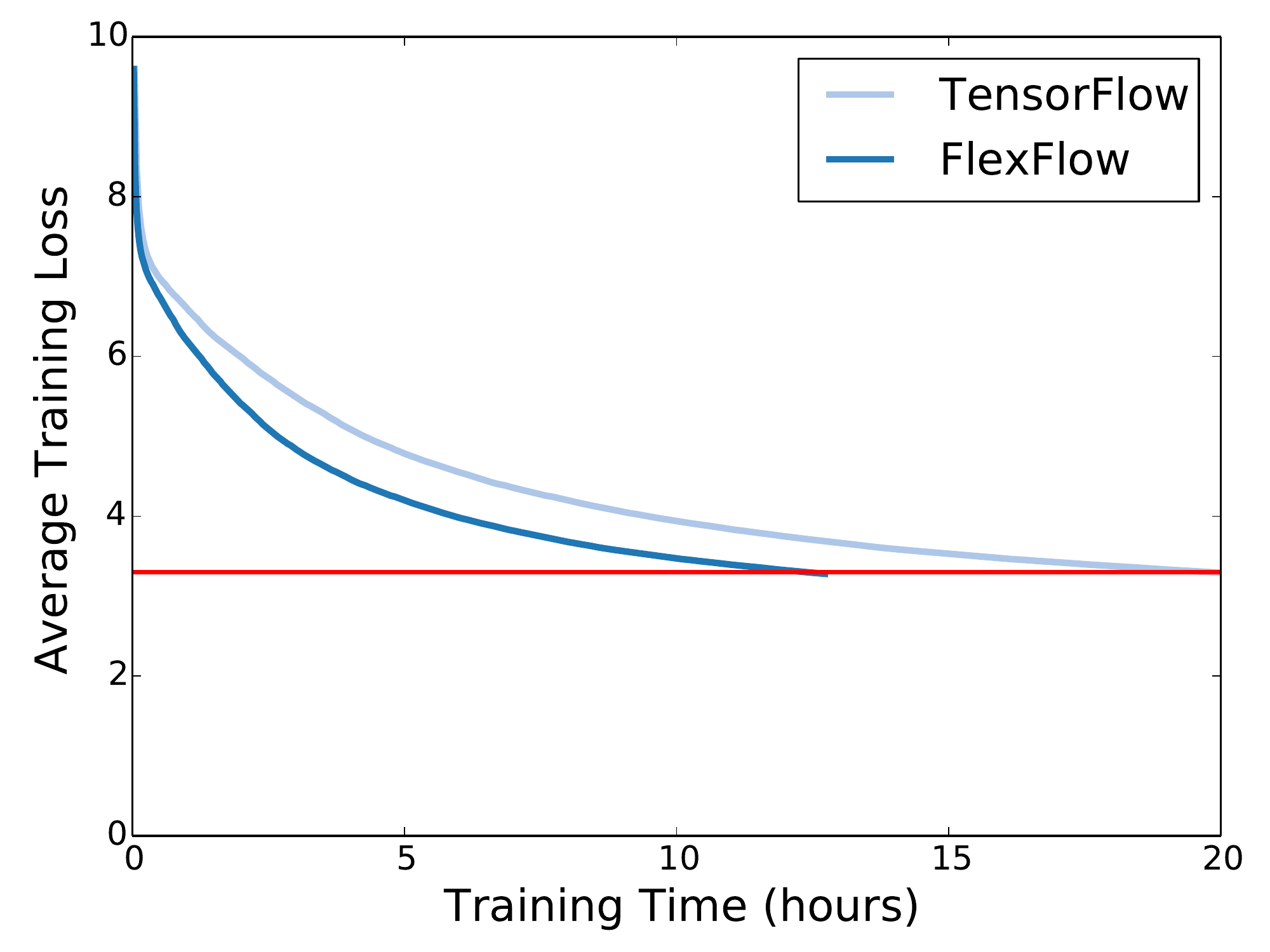}
\vspace{-3mm}
\caption{Training curves of \inception in different systems. 
The model is trained on 16 P100 GPUs (4 nodes).}
\vspace{-3mm}
\label{fig:end_to_end_inception}
\end{figure}
\Sys performs the same computation as other deep learning systems for a DNN model and therefore achieves the same model accuracy. 
Table~\ref{tab:dnns} verifies that \Sys achieves the state-of-the-art accuracies on the DNN benchmarks used in the experiments.

In this experiment, we compare the end-to-end training performance between \Sys and TensorFlow on \inception.
We train \inception on the ImageNet dataset until the model reaches the single-crop top-1 accuracy of 72\% on the validation set. 
The training processes in both frameworks use stochastic gradient decent (SGD) with a learning rate of 0.045 and a weight decay of 0.0001.
Figure~\ref{fig:end_to_end_inception} illustrates the training curves of the two systems on \inception and show that \Sys reduces the end-to-end training time by 38\% compared to TensorFlow.

\begin{table*}[t]
\vspace{-3mm}
\centering
\caption{The end-to-end search time with different simulation algorithms (seconds).}
\label{tab:simulate_runtime}
\vspace{-3mm}
\resizebox{\textwidth}{!}{
\begin{tabular}{|r|llc|llc|llc|llc|llc|llc|}
\hline
{\bf Num.} & \multicolumn{3}{c|}{\bf AlexNet} & \multicolumn{3}{c|}{\bf ResNet} & \multicolumn{3}{c|}{\bf Inception} & \multicolumn{3}{c|}{\bf RNNTC} & \multicolumn{3}{c|}{\bf RNNLM} & \multicolumn{3}{c|}{\bf NMT}\\
{\bf GPUs} & Full & Delta & Speedup & Full & Delta & Speedup & Full & Delta & Speedup & Full & Delta & Speedup & Full & Delta & Speedup & Full & Delta & Speedup\\
\hline
4 & 0.11 & 0.04 & {\bf 2.9$\times$} & 1.4 & 0.4 & {\bf 3.2$\times$} & 14 & 4.1 & {\bf 3.4$\times$} & 16 & 7.5 & {\bf 2.2$\times$} & 21 & 9.2 & {\bf 2.3$\times$} & 40 & 16 & {\bf 2.5$\times$}\\
8 & 0.40 & 0.13 & {\bf 3.0$\times$} & 4.5 & 1.4 & {\bf 3.2$\times$} & 66 & 17 & {\bf 3.9$\times$} & 91 & 39 & {\bf 2.3$\times$} & 76 & 31 & {\bf 2.5$\times$} & 178 & 65 & {\bf 2.7$\times$}\\
16 & 1.4 & 0.48 & {\bf 2.9$\times$} & 22 & 7.3 & {\bf 3.1$\times$} & 388 & 77 & {\bf 5.0$\times$} & 404 & 170 & {\bf 2.4$\times$} & 327 & 121 & {\bf 2.7$\times$} & 998 & 328 & {\bf 3.0$\times$}\\
32 & 5.3 & 1.8 & {\bf 3.0$\times$} & 107 & 33 & {\bf 3.2$\times$} & 1746 & 298 & {\bf 5.9$\times$} & 1358 & 516 & {\bf 2.6$\times$} & 1102 & 342 & {\bf 3.2$\times$} & 2698 & 701 & {\bf 3.8$\times$}\\
64 & 18 & 5.9 & {\bf 3.0$\times$} & 515 & 158 & {\bf 3.3$\times$} & 8817 & 1278 & {\bf 6.9$\times$} & 4404 & 1489 & {\bf 3.0$\times$} & 3406 & 969 & {\bf 3.6$\times$} & 8982 & 2190 & {\bf 4.1$\times$}\\
\hline
\end{tabular}
}
\vspace{-3mm}
\end{table*}

\subsubsection{Automated Parallelization Optimizer}
\begin{figure}
\vspace{-3mm}
\centering
\subfloat[REINFORCE] {
\label{fig:compare_reinforce}
\includegraphics[scale=0.19]{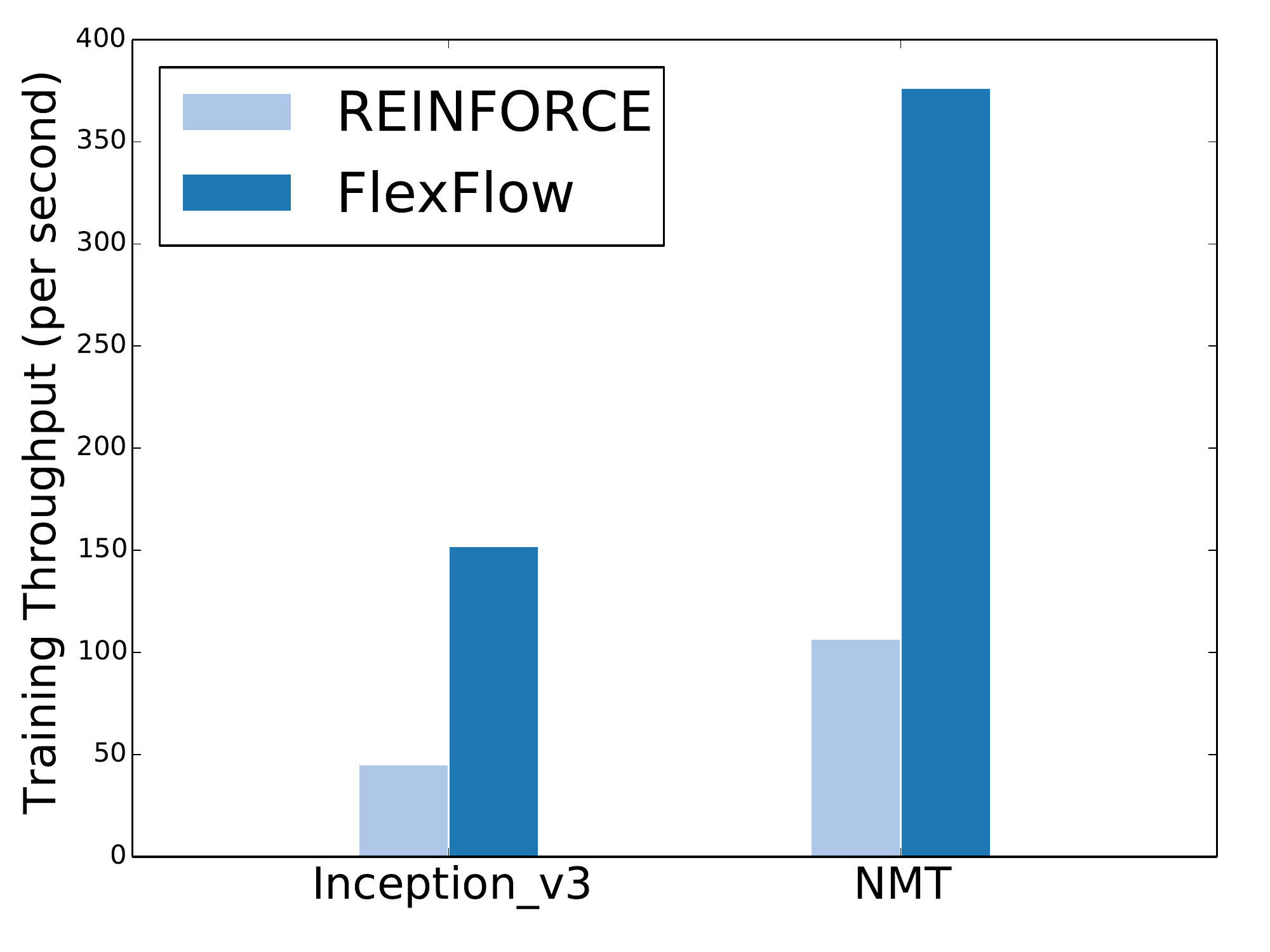}
}
\subfloat[OptCNN] {
\label{fig:compare_optcnn}
\includegraphics[scale=0.19]{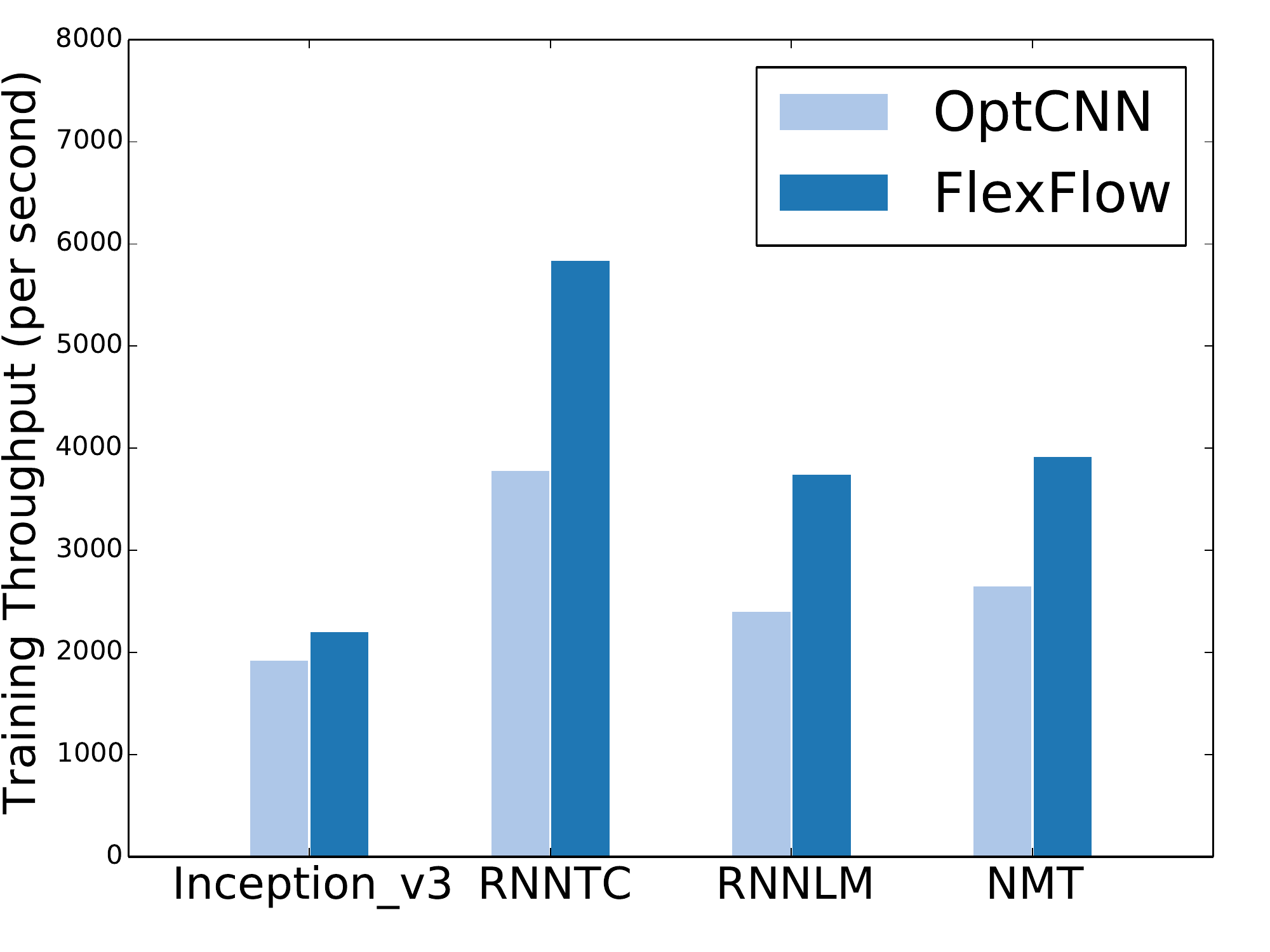}
}
\vspace{-3mm}
\caption{Comparison among the parallelization strategies found by different automated frameworks.}
\label{fig:compare_others}
\vspace{-3mm}
\end{figure}
We compare against two automated frameworks that find parallelization strategies in a limited search space.

{\bf REINFORCE}~\cite{DevicePlace} uses reinforcement learning to learn device placement for model parallelism.
We are not aware of any publicly available implementation of REINFORCE, so we compare against the learned device placement for \inception and NMT, as reported in~\cite{DevicePlace}.

Figure~\ref{fig:compare_reinforce} compares the training throughput of the strategies found by \Sys and REINFORCE for four K80 GPUs on a single node.
The parallelization strategies found by \Sys achieve 3.4 - 3.8$\times$ speedup compared to REINFORCE. We attribute the performance improvement to the larger search space explored by \Sys. 

Besides improving training performance, \Sys has two additional advantages over REINFORCE.
First, REINFORCE requires executing each strategy in the hardware environment to get reward signals and takes 12-27 hours to find the best placement~\cite{DevicePlace}, while the \Sys execution optimizer finds efficient parallelization strategies for these executions in 14-40 seconds.
Second, REINFORCE uses up to 160 compute nodes (with 4 GPUs on each node) to find the placement in time, while \Sys uses a single compute node to run the execution optimizer.

{\bf OptCNN}~\cite{OptCNN} optimizes parallelization for DNNs with linear operator graphs.
OptCNN assumes that different operations in an operator graph cannot be performed in parallel and estimates a DNN's execution time as the sum of the operations' computation time and synchronization time and the tensors' data transfer time. 
This assumption allows OptCNN to use a dynamic programming algorithm to find an efficient parallelization strategy.

We compare the strategies found by \Sys and OptCNN for different DNNs on 16 P100 GPUs. 
The frameworks found the same parallelization strategies for AlexNet and ResNet with linear operator graphs and found different strategies for the other DNNs as shown in Figure~\ref{fig:compare_optcnn}.
For these DNNs with non-linear operator graphs, \Sys achieves 1.2-1.6$\times$ speedup compared to OptCNN by using parallelization strategies that exploit parallelism across different operations. We show two examples in Section~\ref{subsec:case_study}.


\subsection{Execution Simulator}
\label{subsec:eval_simulator}
\begin{figure}
\vspace{-4mm}
\centering
\subfloat[\inception] {
\label{fig:simulator_acc_inception}
\includegraphics[scale=0.19]{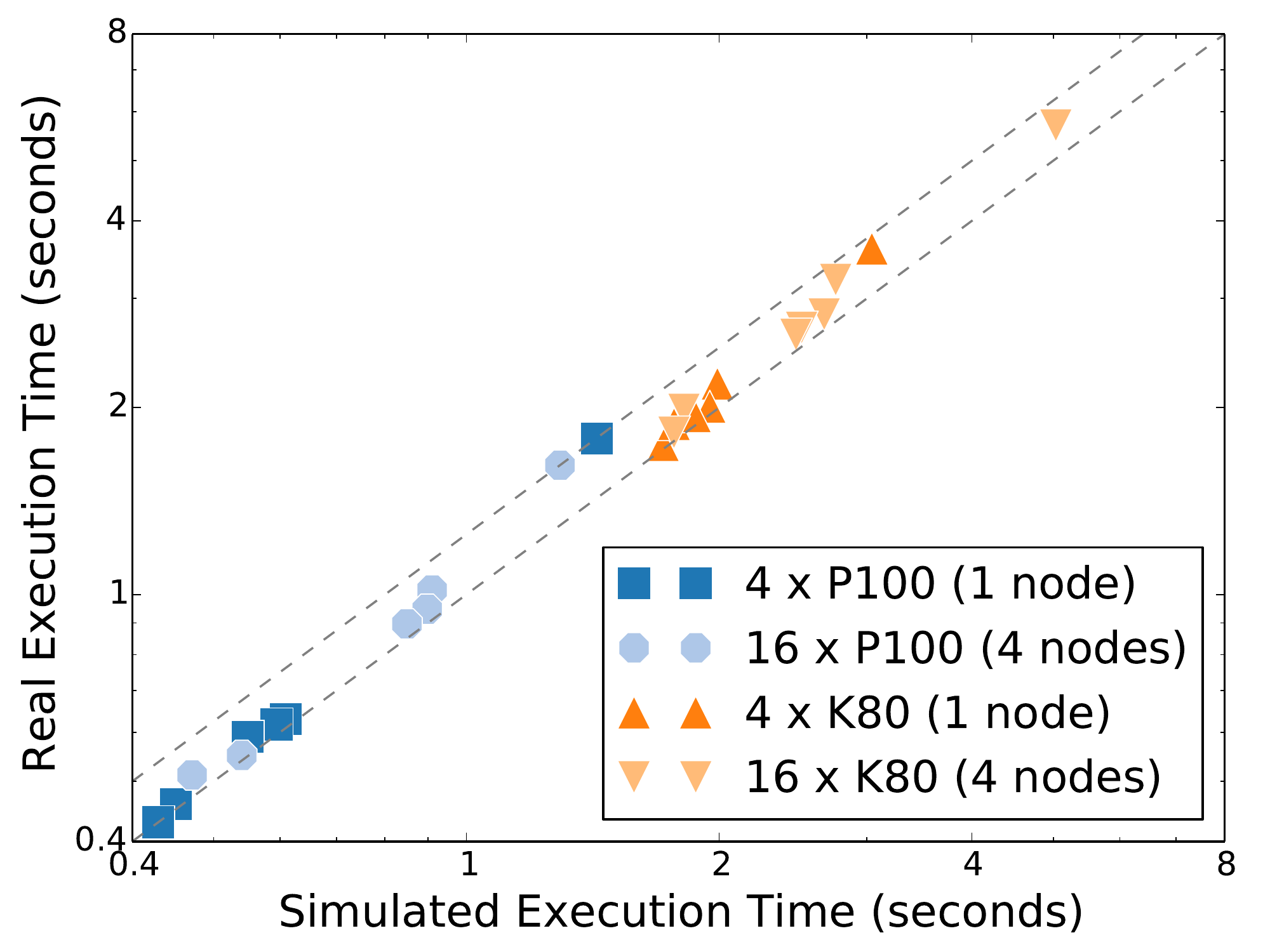}
}
\subfloat[NMT] {
\label{fig:simulator_acc_nmt}
\includegraphics[scale=0.19]{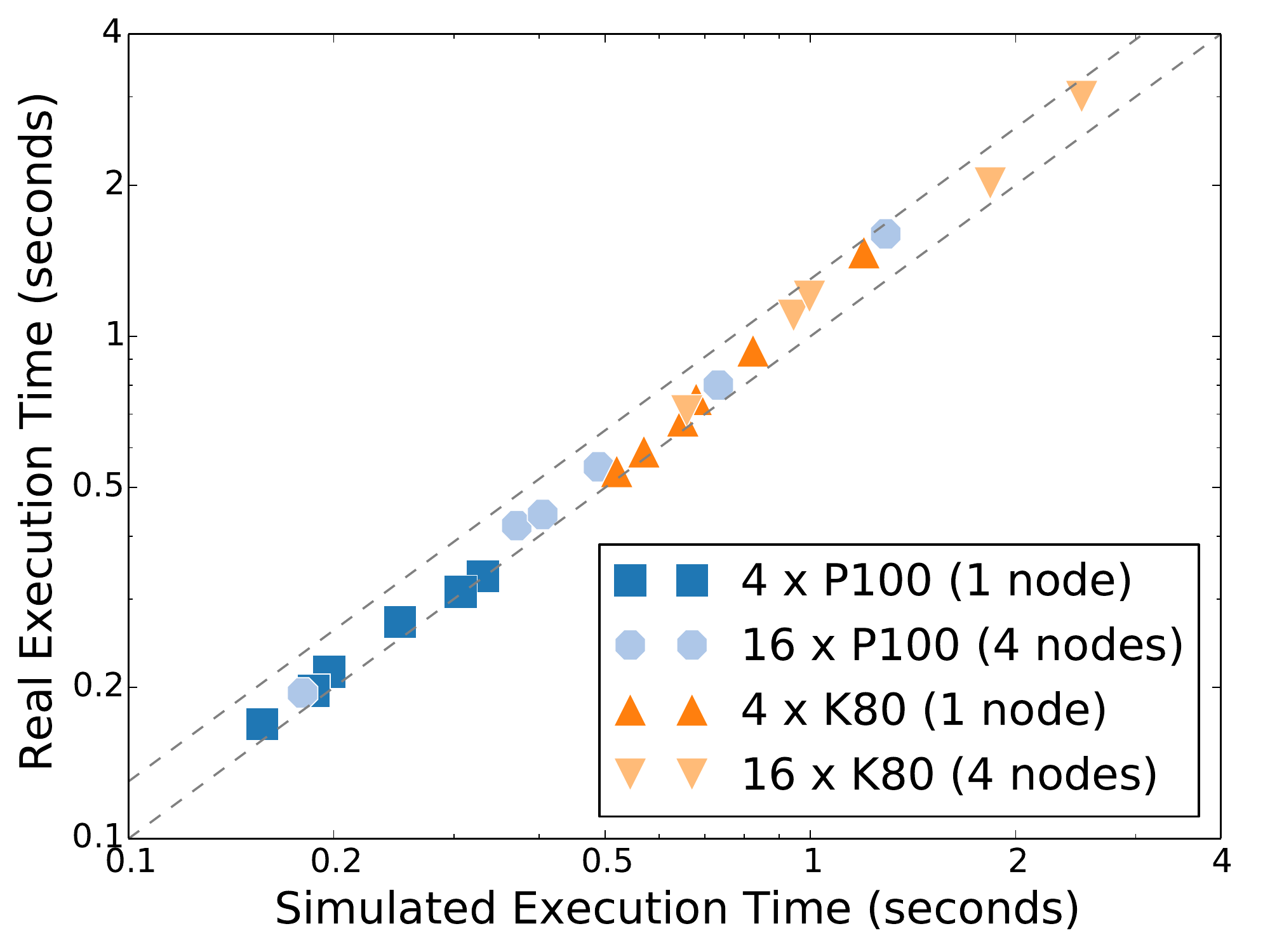}
}
\vspace{-3mm}
\caption{Comparison between the simulated and actual execution time for different DNNs and device topologies.}
\vspace{-3mm}
\label{fig:simulator_acc}
\end{figure}
We evaluate the performance of the simulator using two metrics: simulator accuracy and simulator execution time.
\subsubsection{Simulator Accuracy} 
In this experiment, we compare the estimated execution time predicted by the execution simulator with the real execution time measured by actual executions. 
Figure~\ref{fig:simulator_acc} shows the results for different DNNs and different available devices.
The dashed lines indicate a relative difference of 0\% and 30\%, respectively, which encompasses the variance between actual and predicted execution time.
In addition, for different parallelization strategies with the same operator graph and device topology (i.e., points of the same shape in the figure), their simulated execution time preserves actual execution time ordering, which shows that simulated execution time is an appropriate metric to evaluate the performance of different strategies.

\subsubsection{Simulator Execution Time}
\label{subsec:simulator_execution_time}
\begin{figure}
\centering
\includegraphics[scale=0.2]{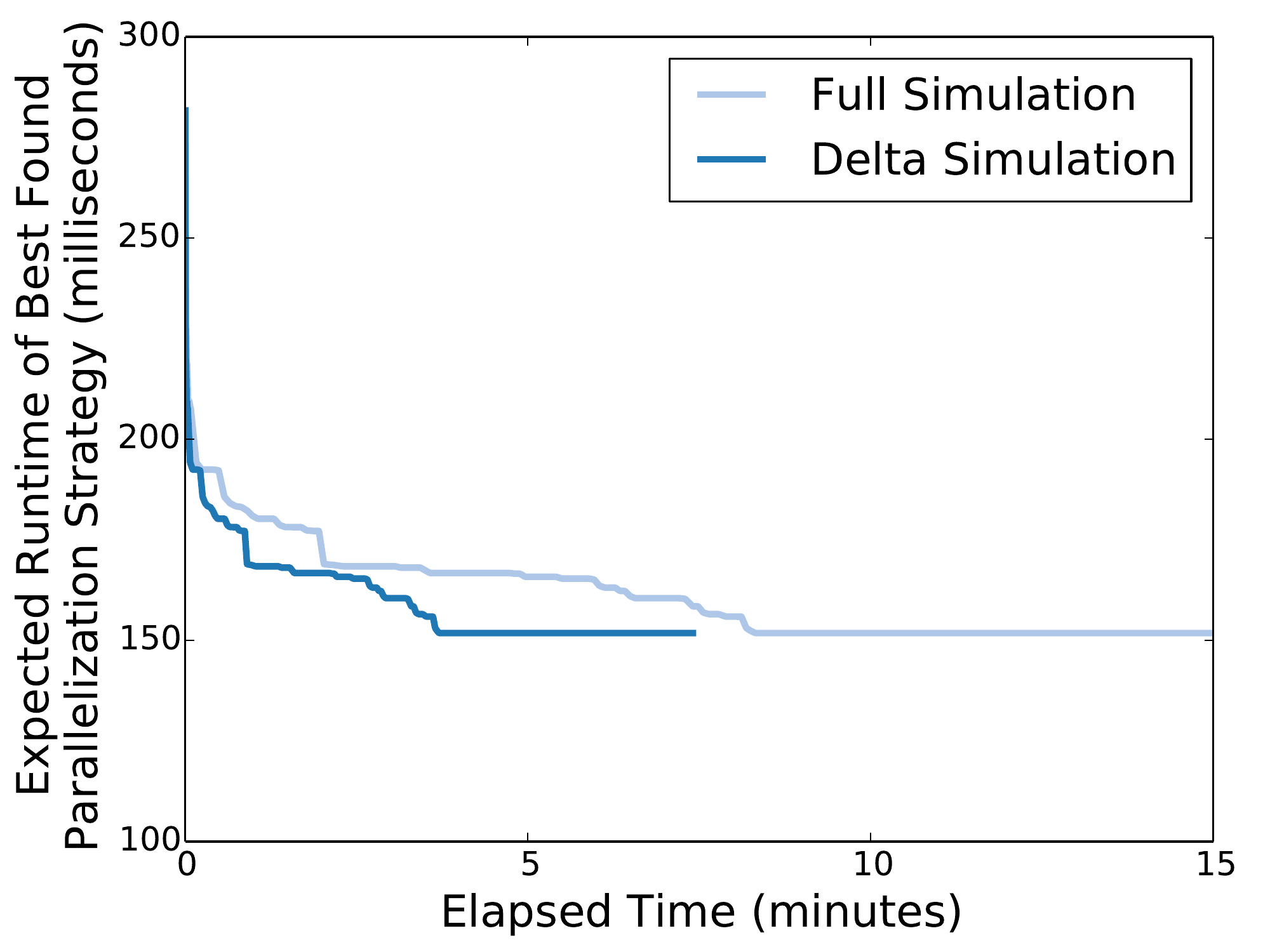}
\vspace{-3mm}
\caption{Search performance with the full and delta simulation algorithms for the NMT model on 16 P100 GPUs (4 nodes).}
\vspace{-3mm}
\label{fig:simulate_timeline}
\end{figure}
Figure~\ref{fig:simulate_timeline} shows the search performance with different simulation algorithms for finding a strategy for the NMT model on 16 P100 GPUs on 4 nodes.
The full and delta simulation algorithms terminate in 16 and 6 minutes, respectively.
If the allowed time budget is less than 8 minutes, the full simulation algorithm will find a worse strategy than the delta simulation algorithm.

We compare the end-to-end search time of the execution optimizer with different simulation algorithms.
For a given DNN model and device topology, we measure the average execution time of the optimizer using 10 random initial strategies.
The results are shown in Table~\ref{tab:simulate_runtime}. The delta simulation algorithm is 2.2-6.9$\times$ faster than the full simulation algorithm.
Moreover, the speedup over the full simulation algorithm increases as we scale the number of devices.

\subsection{Search Algorithm}
\label{subsec:eval_optimizer}
\begin{figure*}[t!]
\vspace{-7mm}
\centering
\includegraphics[scale=0.4]{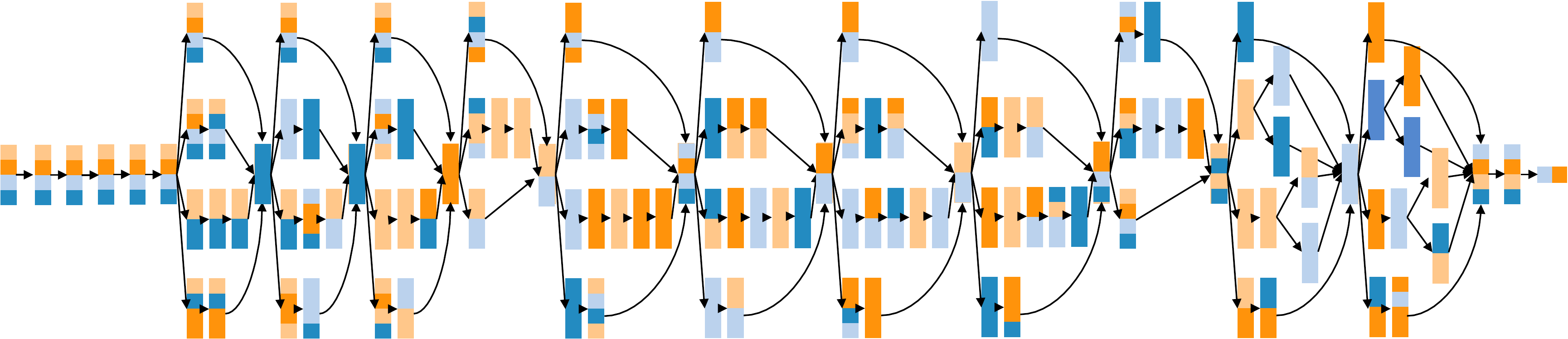}
\vspace{-3mm}
\caption{The best strategy for parallelizing the \inception model on 4 P100 GPUs. 
For each operation, the vertical and horizontal dimensions indicate parallelism in the batch and channel dimension, respectively. Each GPU is denoted by a color. 
This strategy reduces the per-iteration execution time by 12\% compared to data parallelism.}
\vspace{-5mm}
\label{fig:case_study_inception}
\end{figure*}
This section evaluates the quality of the best parallelization strategies discovered by the search algorithm.

First, we compare the best discovered strategies with the global optimal strategies for small executions.
To obtain a search space of reasonable size, we limit the number of devices to 4 and consider the following two DNNs.
LeNet~\cite{lenet} is a 6-layer CNN for image classification. The second DNN is a variant of RNNLM where the number of unrolling steps for each recurrent layer is restricted to 2.
The search space for both DNNs contains approximately $10^{11}$ strategies.
We use depth-first search to explore the search space and use A$^*$~\cite{IntroAlg} to prune the search space. 
Finding the optimal strategies for LeNet and RNNLM took 0.8 and 18 hours, respectively. For both DNNs, \Sys finds the global optimal strategy.

Second, we test if the search algorithm returns at least a locally optimal strategy in larger search spaces by comparing the best discovered strategy with all of its neighbors.
For this experiment, we consider all six DNNs on 2, 4, and 8 devices, where the number of neighbors remains small enough to exhaustively enumerate them all. 
All the strategies returned by FlexFlow were locally optimal.

\subsection{Case Studies}
\label{subsec:case_study}
\vspace{-0mm}
\begin{figure}
\centering
\includegraphics[scale=0.23]{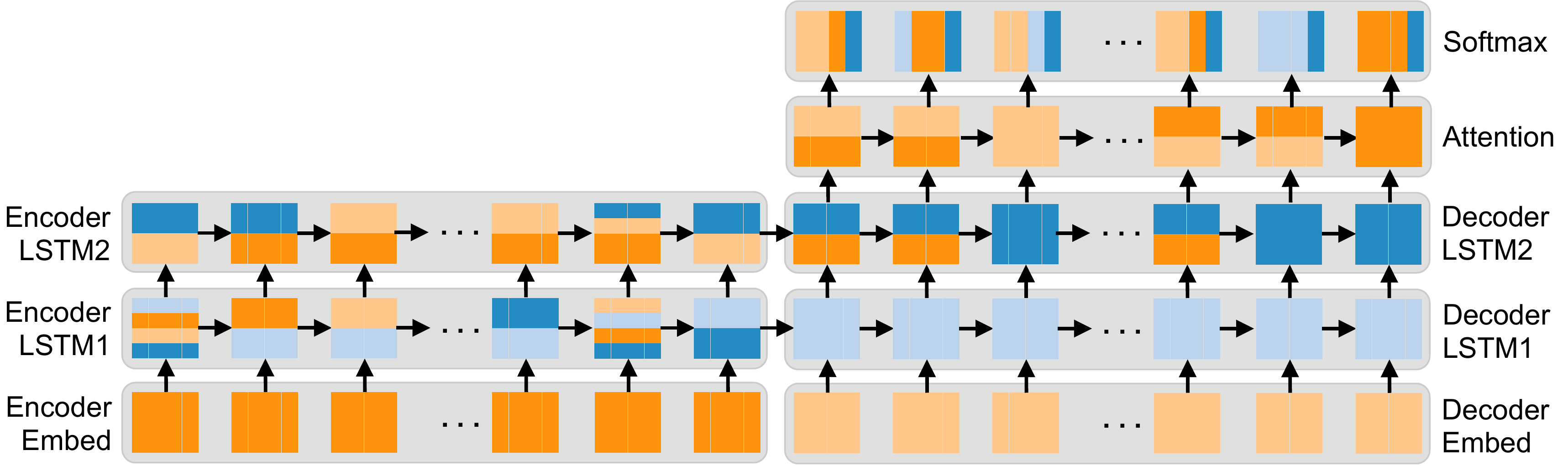}
\vspace{-7mm}
\caption{The best strategy for parallelizing the NMT model on 4 P100 GPUs. For each operation, the vertical and horizontal dimensions indicate parallelism in the batch and channel dimension, respectively. Each grey box denotes a layer, whose operations share the same network parameters. Each GPU is denoted by a color.}
\label{fig:case_study_nmt}
\vspace{-4mm}
\end{figure}


We discuss the best strategies discovered by \Sys and how they improve parallelization performance.

{\bf \inception.} Figure~\ref{fig:case_study_inception} shows the best discovered strategy for parallelizing \inception on four P100 GPUs on a single node, which exploits intra-operation parallelism for operations on the critical path and uses a combination of intra- and inter-operation parallelism for operations on different branches. 
This results in a well-balanced workload and reduces data transfers for parameter synchronization. 
Compared to data parallelism, this strategy reduces the parameter synchronization costs by 75\% and the per-iteration execution time by 12\%.

For parallelizing the same \inception model on four K80 GPUs with asymmetric connections between GPUs (see Figure~\ref{fig:xstream_arch}), we observe that the best discovered strategy tends to parallelize operations on adjacent GPUs with a direct connection to reduce the communication costs.

{\bf NMT.} Figure~\ref{fig:case_study_nmt} shows the best discovered strategy for parallelizing NMT on four P100 GPUs, which uses various strategies for parallelizing different layers. 
We briefly discuss the insights from this strategy. 
First, for a layer with a large number of network parameters and little computation (e.g., the embed layer), it is beneficial to perform the computation on a small number of GPU devices to reduce parameter synchronization costs.
Second, for a layer with a large number of network parameters and a heavy computation workload (e.g., the softmax layer), \Sys uses parallelism in the channel dimension and assigns the computation for a subset of channels to each task. 
This allows each device to use a subset of the network parameters, which reduces parameter synchronization costs while maintaining load balance. 
Third, for multiple recurrent layers (e.g., the LSTM and attention layers), \Sys uses concurrency among different layers as well as parallelism within each operation to cooperatively reduce parameter synchronization costs while balancing load. 

\section{Conclusion}
This paper presents \Sys, a deep learning system that automatically finds efficient parallelization strategies for DNN applications.
\Sys uses a guided randomized search procedure to explore the space of possible strategies and includes an execution simulator that is an efficient and accurate predictor of DNN performance.
We evaluate \Sys with six real-world DNN benchmarks on two GPU clusters and show \Sys significantly outperforms state-of-the-art parallelization approaches.

\bibliography{bibliography}

\begin{thebibliography}{10}

\bibitem{moviereviews}
Movie review data.
\newblock \url{https://www.cs.cornell.edu/people/pabo/movie-review-data/},
  2005.

\bibitem{caffe2}
{A New Lightweight, Modular, and Scalable Deep Learning Framework}.
\newblock \url{https://caffe2.ai}, 2016.

\bibitem{wmt16}
Conference on machine translation.
\newblock \url{http://www.statmt.org/wmt16}, 2016.

\bibitem{cublas}
{Dense Linear Algebra on GPUs}.
\newblock \url{https://developer.nvidia.com/cublas}, 2016.

\bibitem{tensorflowbenchmark}
{TensorFlow Benchmarks}.
\newblock \url{https://www.tensorflow.org/performance/benchmarks}, 2017.

\bibitem{pytorch}
{Tensors and Dynamic neural networks in Python with strong GPU acceleration.}
\newblock \url{https://pytorch.org}, 2017.

\bibitem{tensorflow}
M.~Abadi, P.~Barham, J.~Chen, Z.~Chen, A.~Davis, J.~Dean, M.~Devin,
  S.~Ghemawat, G.~Irving, M.~Isard, M.~Kudlur, J.~Levenberg, R.~Monga,
  S.~Moore, D.~G. Murray, B.~Steiner, P.~Tucker, V.~Vasudevan, P.~Warden,
  M.~Wicke, Y.~Yu, and X.~Zheng.
\newblock Tensorflow: A system for large-scale machine learning.
\newblock In {\em Proceedings of the 12th USENIX Conference on Operating
  Systems Design and Implementation}, OSDI, 2016.

\bibitem{speech2}
D.~Amodei, S.~Ananthanarayanan, R.~Anubhai, J.~Bai, E.~Battenberg, C.~Case,
  J.~Casper, B.~Catanzaro, Q.~Cheng, G.~Chen, J.~Chen, J.~Chen, Z.~Chen,
  M.~Chrzanowski, A.~Coates, G.~Diamos, K.~Ding, N.~Du, E.~Elsen, J.~Engel,
  W.~Fang, L.~Fan, C.~Fougner, L.~Gao, C.~Gong, A.~Hannun, T.~Han, L.~V.
  Johannes, B.~Jiang, C.~Ju, B.~Jun, P.~LeGresley, L.~Lin, J.~Liu, Y.~Liu,
  W.~Li, X.~Li, D.~Ma, S.~Narang, A.~Ng, S.~Ozair, Y.~Peng, R.~Prenger,
  S.~Qian, Z.~Quan, J.~Raiman, V.~Rao, S.~Satheesh, D.~Seetapun, S.~Sengupta,
  K.~Srinet, A.~Sriram, H.~Tang, L.~Tang, C.~Wang, J.~Wang, K.~Wang, Y.~Wang,
  Z.~Wang, Z.~Wang, S.~Wu, L.~Wei, B.~Xiao, W.~Xie, Y.~Xie, D.~Yogatama,
  B.~Yuan, J.~Zhan, and Z.~Zhu.
\newblock Deep speech 2: End-to-end speech recognition in english and mandarin.
\newblock In {\em Proceedings of the 33rd International Conference on
  International Conference on Machine Learning}, ICML'16.

\bibitem{NMT2}
D.~Bahdanau, K.~Cho, and Y.~Bengio.
\newblock Neural machine translation by jointly learning to align and
  translate.
\newblock {\em CoRR}, abs/1409.0473, 2014.

\bibitem{Legion12}
M.~Bauer, S.~Treichler, E.~Slaughter, and A.~Aiken.
\newblock Legion: Expressing locality and independence with logical regions.
\newblock In {\em Proceedings of the International Conference on High
  Performance Computing, Networking, Storage and Analysis}, 2012.

\bibitem{MSGBK}
C.~Chelba, T.~Mikolov, M.~Schuster, Q.~Ge, T.~Brants, and P.~Koehn.
\newblock One billion word benchmark for measuring progress in statistical
  language modeling.
\newblock {\em CoRR}, abs/1312.3005, 2013.

\bibitem{mxnet}
T.~Chen, M.~Li, Y.~Li, M.~Lin, N.~Wang, M.~Wang, T.~Xiao, B.~Xu, C.~Zhang, and
  Z.~Zhang.
\newblock {MXNet}: {A} flexible and efficient machine learning library for
  heterogeneous distributed systems.
\newblock {\em CoRR}, abs/1512.01274, 2015.

\bibitem{cudnn}
S.~Chetlur, C.~Woolley, P.~Vandermersch, J.~Cohen, J.~Tran, B.~Catanzaro, and
  E.~Shelhamer.
\newblock cudnn: Efficient primitives for deep learning.
\newblock {\em CoRR}, abs/1410.0759, 2014.

\bibitem{IntroAlg}
T.~H. Cormen, C.~E. Leiserson, R.~L. Rivest, and C.~Stein.
\newblock {\em Introduction to Algorithms, Third Edition}.
\newblock The MIT Press, 3rd edition, 2009.

\bibitem{distbelief}
J.~Dean, G.~S. Corrado, R.~Monga, K.~Chen, M.~Devin, Q.~V. Le, M.~Z. Mao,
  M.~Ranzato, A.~Senior, P.~Tucker, K.~Yang, and A.~Y. Ng.
\newblock Large scale distributed deep networks.
\newblock In {\em NIPS}, 2012.

\bibitem{imagenet2009}
J.~Deng, W.~Dong, R.~Socher, L.-J. Li, K.~Li, and L.~Fei-Fei.
\newblock {ImageNet}: A large-scale hierarchical image database.
\newblock In {\em Proceedings of the IEEE Conference on Computer Vision and
  Pattern Recognition}, CVPR, 2009.

\bibitem{MCMC}
W.~R. Gilks, S.~Richardson, and D.~Spiegelhalter.
\newblock {\em Markov chain Monte Carlo in practice}.
\newblock CRC press, 1995.

\bibitem{firmament}
I.~Gog, M.~Schwarzkopf, A.~Gleave, R.~N.~M. Watson, and S.~Hand.
\newblock Firmament: Fast, centralized cluster scheduling at scale.
\newblock In {\em 12th {USENIX} Symposium on Operating Systems Design and
  Implementation ({OSDI} 16)}, pages 99--115, Savannah, GA, 2016. {USENIX}
  Association.

\bibitem{LargeSGD}
P.~Goyal, P.~Doll{\'{a}}r, R.~B. Girshick, P.~Noordhuis, L.~Wesolowski,
  A.~Kyrola, A.~Tulloch, Y.~Jia, and K.~He.
\newblock Accurate, large minibatch {SGD:} training imagenet in 1 hour.
\newblock {\em CoRR}, abs/1706.02677, 2017.

\bibitem{speech1}
A.~Graves and N.~Jaitly.
\newblock Towards end-to-end speech recognition with recurrent neural networks.
\newblock In {\em Proceedings of the 31st International Conference on
  International Conference on Machine Learning}, ICML'14, 2014.

\bibitem{MetropolisHastings}
W.~K. Hastings.
\newblock Monte carlo sampling methods using markov chains and their
  applications.
\newblock {\em Biometrika}, 57(1):97--109, 1970.

\bibitem{resnet}
K.~He, X.~Zhang, S.~Ren, and J.~Sun.
\newblock Deep residual learning for image recognition.
\newblock In {\em Proceedings of the IEEE Conference on Computer Vision and
  Pattern Recognition}, CVPR, 2016.

\bibitem{densenet}
G.~Huang, Z.~Liu, and K.~Q. Weinberger.
\newblock Densely connected convolutional networks.
\newblock {\em CoRR}, abs/1608.06993, 2016.

\bibitem{quincy}
M.~Isard, V.~Prabhakaran, J.~Currey, U.~Wieder, K.~Talwar, and A.~Goldberg.
\newblock Quincy: Fair scheduling for distributed computing clusters.
\newblock In {\em Proceedings of the ACM SIGOPS 22nd Symposium on Operating
  Systems Principles}, SOSP '09, pages 261--276. ACM, 2009.

\bibitem{OptCNN}
Z.~Jia, S.~Lin, C.~R. Qi, and A.~Aiken.
\newblock Exploring hidden dimensions in parallelizing convolutional neural
  networks.
\newblock {\em CoRR}, abs/1802.04924, 2018.

\bibitem{RNNTC}
Y.~Kim.
\newblock Convolutional neural networks for sentence classification.
\newblock {\em CoRR}, abs/1408.5882, 2014.

\bibitem{OWT}
A.~Krizhevsky.
\newblock One weird trick for parallelizing convolutional neural networks.
\newblock {\em CoRR}, abs/1404.5997, 2014.

\bibitem{alexnet}
A.~Krizhevsky, I.~Sutskever, and G.~E. Hinton.
\newblock {ImageNet} classification with deep convolutional neural networks.
\newblock In {\em Proceedings of the 25th International Conference on Neural
  Information Processing Systems}, NIPS, 2012.

\bibitem{lam1977worst}
S.~Lam and R.~Sethi.
\newblock Worst case analysis of two scheduling algorithms.
\newblock {\em SIAM Journal on Computing}, 6, 1977.

\bibitem{lenet}
Y.~LeCun.
\newblock {LeNet-5}, convolutional neural networks.
\newblock {\em URL: http://yann. lecun. com/exdb/lenet}, 2015.

\bibitem{penntreebank}
M.~P. Marcus, M.~A. Marcinkiewicz, and B.~Santorini.
\newblock Building a large annotated corpus of english: The penn treebank.
\newblock {\em Comput. Linguist.}, 19.

\bibitem{DevicePlace2}
A.~Mirhoseini, A.~Goldie, H.~Pham, B.~Steiner, Q.~V. Le, and J.~Dean.
\newblock A hierarchical model for device placement.
\newblock In {\em International Conference on Learning Representations}, 2018.

\bibitem{DevicePlace}
A.~Mirhoseini, H.~Pham, Q.~V. Le, B.~Steiner, R.~Larsen, Y.~Zhou, N.~Kumar,
  M.~Norouzi, S.~Bengio, and J.~Dean.
\newblock Device placement optimization with reinforcement learning.
\newblock 2017.

\bibitem{bleu}
K.~Papineni, S.~Roukos, T.~Ward, and W.-J. Zhu.
\newblock Bleu: A method for automatic evaluation of machine translation.
\newblock In {\em Proceedings of the 40th Annual Meeting on Association for
  Computational Linguistics}, ACL '02, 2002.

\bibitem{ILSVRC15}
O.~Russakovsky, J.~Deng, H.~Su, J.~Krause, S.~Satheesh, S.~Ma, Z.~Huang,
  A.~Karpathy, A.~Khosla, M.~Bernstein, A.~C. Berg, and L.~Fei-Fei.
\newblock {ImageNet Large Scale Visual Recognition Challenge}.
\newblock {\em International Journal of Computer Vision (IJCV)},
  115(3):211--252, 2015.

\bibitem{imagenet}
O.~Russakovsky, J.~Deng, H.~Su, J.~Krause, S.~Satheesh, S.~Ma, Z.~Huang,
  A.~Karpathy, A.~Khosla, M.~Bernstein, et~al.
\newblock Imagenet large scale visual recognition challenge.
\newblock {\em International Journal of Computer Vision}, 2015.

\bibitem{game}
D.~Silver, A.~Huang, C.~J. Maddison, A.~Guez, L.~Sifre, G.~Van Den~Driessche,
  J.~Schrittwieser, I.~Antonoglou, V.~Panneershelvam, M.~Lanctot, et~al.
\newblock Mastering the game of go with deep neural networks and tree search.
\newblock {\em Nature}, 529:484--489, 2016.

\bibitem{vgg}
K.~Simonyan and A.~Zisserman.
\newblock Very deep convolutional networks for large-scale image recognition.
\newblock {\em CoRR}, abs/1409.1556, 2014.

\bibitem{GoogleNet}
C.~Szegedy, W.~Liu, Y.~Jia, P.~Sermanet, S.~E. Reed, D.~Anguelov, D.~Erhan,
  V.~Vanhoucke, and A.~Rabinovich.
\newblock Going deeper with convolutions.
\newblock {\em CoRR}, abs/1409.4842, 2014.

\bibitem{inception}
C.~Szegedy, V.~Vanhoucke, S.~Ioffe, J.~Shlens, and Z.~Wojna.
\newblock Rethinking the inception architecture for computer vision.
\newblock In {\em Proceedings of the IEEE Conference on Computer Vision and
  Pattern Recognition}, 2016.

\bibitem{dependentpartitioning}
S.~Treichler, M.~Bauer, R.~Sharma, E.~Slaughter, and A.~Aiken.
\newblock Dependent partitioning.
\newblock In {\em Proceedings of the 2016 ACM SIGPLAN International Conference
  on Object-Oriented Programming, Systems, Languages, and Applications},
  OOPSLA' 16. ACM, 2016.

\bibitem{GNMT}
Y.~Wu, M.~Schuster, Z.~Chen, Q.~V. Le, M.~Norouzi, W.~Macherey, M.~Krikun,
  Y.~Cao, Q.~Gao, K.~Macherey, J.~Klingner, A.~Shah, M.~Johnson, X.~Liu,
  L.~Kaiser, S.~Gouws, Y.~Kato, T.~Kudo, H.~Kazawa, K.~Stevens, G.~Kurian,
  N.~Patil, W.~Wang, C.~Young, J.~Smith, J.~Riesa, A.~Rudnick, O.~Vinyals,
  G.~Corrado, M.~Hughes, and J.~Dean.
\newblock Google's neural machine translation system: Bridging the gap between
  human and machine translation.
\newblock {\em CoRR}, abs/1609.08144, 2016.

\bibitem{RNNLM}
W.~Zaremba, I.~Sutskever, and O.~Vinyals.
\newblock Recurrent neural network regularization.
\newblock {\em CoRR}, abs/1409.2329, 2014.

\end{thebibliography}
\bibliographystyle{abbrv}

\end{document}